\documentclass[a4paper,fleqn,usenatbib]{mnras}
\usepackage{mathptmx}
\usepackage{txfonts}
\usepackage[T1]{fontenc}
\usepackage{ae,aecompl}
\usepackage{graphicx}

\title[DYNAMO-{\em HST} Survey]{DYNAMO-HST Survey: Clumps in Nearby Massive Turbulent
  Disks and the Effects of Clump Clustering on Kiloparsec Scale Measurements
  of Clumps}

\author[D.B.~Fisher et al.]{
David~B.~Fisher$^1$\thanks{email:dfisher@swin.edu.au},
Karl Glazebrook$^{1,2}$,
Ivana Damjanov$^8$,
Roberto G. Abraham$^7$,
\newauthor Danail Obreschkow$^{2,4}$,
Emily Wisnioski$^6$,
Robert Bassett$^1$,
Andy Green$^9$, and
\newauthor Peter McGregor$^{10}$
\\
$^1$Centre for Astrophysics and Supercomputing, Swinburne
  University of Technology, P.O. Box 218, Hawthorn, VIC 3122,
  Australia \\
$^2$ARC Centre of
  Excellence for All-sky Astrophysics (CAASTRO) \\
$^4$International Centre for Radio Astronomy Research (ICRAR),
  M468, University of Western Australia, 35 Stirling Hwy, Crawley, \\
  WA 6009, Australia\\
$^6$Max-Planck-Institut f\"ur extraterrestrische Physik, Postfach
  1312, Giessenbachstr., D-85741 Garching, Germany\\
$^7$Department of
  Astronomy \& Astrophysics, University of Toronto, 50 St. George St.,
  Toronto, ON M5S 3H8, Canada\\
 $^8$ Harvard-Smithsonian Center
  for Astrophysics, 60 Garden St., Cambridge, MA 02138,
  USA\\
$^9$Australian Astronomical Observatory, P.O. Box 970,
  North Ryde, NSW 1670, Australia\\
$^{10}$Research School of
  Astronomy and Astrophysics, Australian National University, Cotter
  Rd, Weston, ACT 2611, Australia
}

\date{Accepted August 2016}

\pubyear{2016}

\begin{document}
\label{firstpage}
\pagerange{\pageref{firstpage}--\pageref{lastpage}}
\maketitle

\begin{abstract}
  We present $\sim$100 pc resolution Hubble Space Telescope H$\alpha$
  images of 10 galaxies from the DYnamics of Newly-Assembled Massive
  Objects (DYNAMO) survey of low-$z$ turbulent disk galaxies, and use
  these to undertake the first detailed systematic study of the
  effects of resolution and clump clustering on observations of clumps
  in turbulent disks.  In the DYNAMO-{\em HST} sample we measure clump
  diameters spanning the range $d_{clump} \sim 100-800$~pc, and
  individual clump star formation rates as high as $\sim
  5$~M$_{\odot}$~yr$^{-1}$. DYNAMO clumps have very high SFR surface
  densities, $\Sigma_{SFR}\sim 1-
  15$~M$_{\odot}$~yr$^{-1}$~kpc$^{-2}$, $\sim100\times$ higher than in
  H{\sc ii} regions of nearby spirals. Indeed, SFR surface density provides
  a simple dividing line between massive star forming clumps and local
  star forming regions, where massive star forming clumps have
  $\Sigma_{SFR}> 0.5$~M$_{\odot}$~yr$^{-1}$~kpc$^{-2}$.  When degraded
  to match the observations of galaxies in $z\sim 1-3$ surveys, DYNAMO
  galaxies are similar in morphology and measured clump properties to
  clumpy galaxies observed in the high-$z$ Universe.  Emission peaks
  in the simulated high-redshift maps typically correspond to multiple
  clumps in full resolution images. This clustering of clumps
  systematically increases the apparent size and SFR of clumps in
  1~kpc resolution maps, and decreases the measured SFR surface
  density of clumps by as much as a factor of 20$\times$.  From these
  results we can infer that clump clustering is likely to strongly
  effect the measured properties of clumps in high-$z$ galaxies, which
  commonly have kiloparsec scale resolution.
\end{abstract}


\section{Introduction}

Early Hubble Deep Field observations showed that the morphologies of
typical galaxies in the distant Universe ($z\gtrsim 1$) are more
irregular than those of local galaxies
\citep{abraham1996,conselice2000,elmegreen2005b}.  Subsequent work
\citep[e.g.~][]{elmegreen2005,swinbank2009,genzel2011} has shown that
a large fraction of these so-called ``clumpy'' galaxies are star-forming systems, with
extremely high gas fractions \citep{daddi2010,tacconi2013}. The
clumps themselves correspond to very young stellar populations with
extremely high densities of star formation. Typical clump masses are
$\sim 10^9$~M$_{\odot}$
\citep{elmegreen2005,swinbank2009,genzel2011,wisnioski2012} with star
formation rates up to $\sim10$~M$_{\odot}$~yr$^{-1}$ per clump. The
masses and star formation rates of these clumps are much larger than
those of local star-forming complexes.

One clue to the nature of clumpy galaxies is that a large fraction
of these systems exhibit a
dynamical structure consistent with that of rotating disks whose
gaseous content is turbulent
\citep[eg.~][]{forsterschreiber2009,wuyts2012,wisnioski2011}.  These
two ingredients can lead to large and vigorously star-forming clumps
as the product of dynamical instabilities:
the high gas dispersions generate large Jeans lengths
\citep[e.g.~][]{dekel2009clumps,bournaud2014}, while large gas
fractions fuel very active star formation.

Our ability to test this model is limited by the fact that
observations of star forming regions in high-redshift clumpy galaxies
(using either adaptive optics or the Hubble Space Telescope) are
typically limited to resolutions of $\sim 1$~kpc in the rest frame
\citep[e.g.][]{forster2011a,forster2011b,wuyts2013}, which may be
subject to systematic effects of resolution. Simulations with higher
resolution predict that clumps viewed at kiloparsec resolution likely
are likely composed of several smaller structures, whose sizes are
still larger than the Jeans length for the simulated disk
\citep{behrendt2014arxiv}.  Observationally, finer resolution can be
achieved by targetting gravitationally lensed systems
\citep[e.g.~][]{jones2010,livermore2012,livermore2015,hodge2012},
though uncertainties in lens models limit our ability to interpret
these observations. (For example, strong magnification is only
achieved in a single direction, namely along the critical curve).
Furthermore, it is difficult to construct well-defined samples of
lensed galaxies. In general, lensed samples are biased to galaxies
with smaller masses and lower star formation rates
\citep[e.g.][]{livermore2012,livermore2015}.
  \begin{table*}
	\centering
	\caption{Properties of DYNAMO-{\em HST} sample}
	
	\label{table:sample}
	\begin{tabular}{lcccccccccc}
	\hline
  Galaxy & $z$ & SFR & M$_{star}^a$ & $\sigma_m(H\alpha)^{a,b}$ 
   & $V_{2.2}^a$& Resolution & A$_{H\alpha}^a$ & [NII]/H$\alpha$ & R$_{1/2}$(H$\alpha$) 
   & Morphology$^c$ \\
    &      & M$_{\odot}$ yr$^{-1}$ & $10^{10} M_{\odot}$ & km s$^{-1}$ & km s$^{-1}$ & pc & mag & & kpc &  \\
  G20-2 & 0.1411 & 17.3 $\pm$ 0.7 & 2.16 & 81 & 166 $\pm$
  10 & 90 & 0.83 $\pm$ 0.03 &
  0.46  & 2.1  &  disk  \\
  G13-1 & 0.1388 & 26.5 $\pm$ 0.8 & 1.11 & 76 & 112 $\pm$ 2 & 93 & 0.50
  $\pm$ 0.03 &
  0.19  & 2.6 &  merger  \\
  G14-1 & 0.1323 & 8.3 $\pm$ 0.9 & 2.23 & 70 & 136$\pm$ 8 & 69 & 0.86
  $\pm$ 0.04 &
  0.53  & 1.1 & disk  \\
  G08-5 & 0.1322 & 16.6 $\pm$ 1.0 & 1.73 & 64 & 243 $\pm$ 15 & 108 &
  1.31 $\pm$ 0.06 &
  0.47  & 1.8 &  disk  \\
  H10-2 & 0.1491 & 25.4 $\pm$ 2.7 & 0.95 & 59 & 62 $\pm$ 29 & 117 &
  0.21 $\pm$ 0.04 &
  0.20  & 2.6 & merger  \\
  G04-1 & 0.1298 & 41.6 $\pm$ 2.2 & 6.47 & 50 & 269 $\pm$ 22 & 148 &
  1.55 $\pm$ 0.05 &
  0.51  & 2.7 & disk  \\
  D13-5 & 0.0753 & 21.2 $\pm$ 0.9 & 5.38 & 46 & 192 $\pm$ 2 & 82 &
  1.46 $\pm$ 0.04 &
  0.52  &  2.0 & disk  \\
  D15-3 & 0.0671 & 13.7 $\pm$ 1.0 & 5.42 & 45 & 240 $\pm$ 3 & 51 &
  1.77 $\pm$ 0.08 &
  0.35  &  2.2 & disk  \\
  C13-1 & 0.0788 & 9.9 $\pm$ 0.9 & 3.58 & 29 & 223 $\pm$ 8 &
  58  &  1.05 $\pm$ 0.10  &  0.42  & 4.2 &  disk  \\
  A04-3 & 0.0691 & 3.4 $\pm$ 0.7 & 4.24 & 10 & 218 $\pm$ 2 &
  49  &  1.32 $\pm$ 0.25  &  0.54  & 3.8 &  disk  \\
  \end{tabular}
 $^a$Value taken from Green et al. (2014), methods
   for determining these properties is described there.  \\
 $^b$Errors on $\sigma_m(H\alpha )$ range
   1-3~km~s$^{-1}$\\ 
   $^c$ Classification of ``disks''
   and ``mergers'', is done using both kinematics and stellar surface
   brightness profiles for all DYNAMO-{\em HST} galaxies, we discuss this
   in the Appendix. \\

\end{table*}

Clumpy disks were common at redshifts $z\sim 1-3$ \citep[][and
references therein]{guo2015}.  This redshift range corresponds to the
epoch at which the co-moving star formation rate density in the
Universe was at its peak
\citep{hopkinsbeacom2006,madau2014}. Therefore, during the main epoch
of massive galaxy building, star formation was occurring in a manner
that is not well-represented by the modes of star formation seen in
nearby galaxies. {\em Not well-represented} does not necessarily mean
{\em completely absent}, however, and there is thus considerable
interest in identifying samples of rare local galaxies whose
properties are similar to those of high redshift turbulent disks. Such
systems can be studied at resolutions higher than the $\sim 1$~kpc
rest-frame limit that limits most high-redshift observations.

For example, galaxies selected by rest-frame $UV$ brightness and
compactness, so-called 'Lyman Break Analogue' galaxies (LBA), have
some properties similar to $z>1$ galaxies
\citep{heckman2005}. Recently, \cite{overzier2009} studied the
properties star forming clumps in a sample of LBAs at $z\sim 0.2-0.3$,
finding clump sizes of $\sim 100$~pc.  However, based on both
spatially resolved kinematics \citep{basuzych2009} and {\em HST}
morphologies \citep{overzier2008}, the vast majority of LBAs resemble
the compact dispersion-dominated high-$z$ galaxies found in samples
such as that of \cite{law2009}, rather than the turbulent disk systems
identified in \cite{forsterschreiber2009}, \cite{genzel2011}, and
\cite{wisnioski2011}. Even if dispersion-dominated systems are disks,
they are much more compact that typical massive disks at $z\sim 1-3$
\cite{law2009}. Compact dispersion-dominated systems exhibit star
formation that is often driven by ongoing major mergers
\citep{basuzych2009}, and it is not obvious that the clumps in LBA
galaxies will resemble the clumps in turbulent disks observed at
$z\sim 1-3$.

Recently, \cite{green2010} report the discovery of a sample of
galaxies at $z\sim 0.1$ whose properties closely match those of high
redshift, turbulent, clumpy disk galaxies. Galaxies in the DYNAMO
({\bf DY}namics of {\bf N}ewly-{\bf A}ssembled {\bf M}assive {\bf
  O}bjects) sample have high gas fractions of $f_{gas}\sim 20-40$\% \citep{fisher2014} and H$\alpha$
velocity dispersions 30-80~km~s$^{-1}$ \citep{green2010,green2013,bassett2014} similar
to those of high-redshift turbulent disks. If DYNAMO galaxies do indeed
provide local analogs to the high-$z$ systems, then a number of
interesting general conclusions about turbulent disks can be drawn
from investigations of these local samples. For example,
\cite{obreschkow2015} shows that the specific angular momentum of
turbulent disks is lower than in low-$z$ disks of similar mass,
suggesting that angular momenta may play an important role in driving
dynamical instabilities. Similarly, \cite{bassett2014} uses DYNAMO
galaxies to measure absorption line kinematics in clumpy disks,
showing that the kinematical properties of the stars in these galaxies
are similar to the kinematical properties of the gas.

In the present paper we report on the properties of massive star forming
clumps in 10 DYNAMO galaxies at $z\sim 0.1$ using new Hubble Space Telescope
H$\alpha$ observations. We will compare DYNAMO galaxy morphology to
that of high redshift galaxies, in effort to show that resolution
degraded images appear similar to high-$z$ galaxies. We then
inveistigate the size, luminosity and surface density of clump in
DYNAMO maps. We use clump measurements in full, and resolution
degraded images to make quantitative tests on the effects of
resolution in clumpy galaxies. 
 Throughout this paper,
we assume a concordance cosmology with \hbox{$H_0$ = 67 km\ $s^{-1}$ Mpc$^{-1}$}, $\Omega_M=0.31$,
and $\Omega_\Lambda=0.69$.
\begin{figure}
\begin{center}
\includegraphics[width=0.5\textwidth]{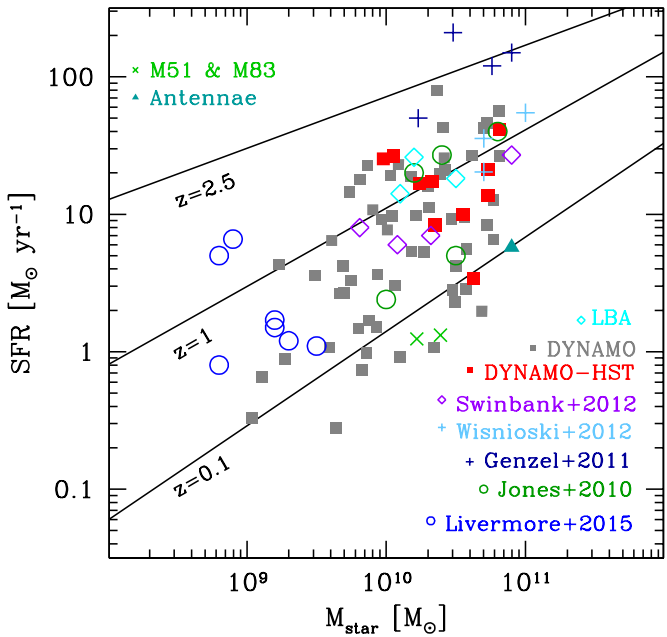}
\end{center}
\caption{ DYNAMO galaxy star formation rates and stellar masses are
  compared to other samples of clumpy galaxies. Grey squares represent
  the full DYNAMO survey \protect\citep{green2013}, and red squares represent
  the subset studied in this paper. The solid lines represent the main sequence of
  star formation for $z=0.1, 1.0, 2.5$ (from bottom to top) taken from
  the analytic fitting function of \protect\cite{whitaker2012}. The ratio of
  stellar mass-to-star formation in DYNAMO galaxies is very similar
  to many of those galaxies in which clumps have previously studied,
  and it is clear that DYNAMO galaxies are very similar to $z\sim 1$
  main sequence galaxies. Details of the comparison galaxy samples
  are described in the Appendix.} 
  \label{fig:ms}
\end{figure}
\begin{figure}
\begin{center}
\includegraphics[width=0.5\textwidth]{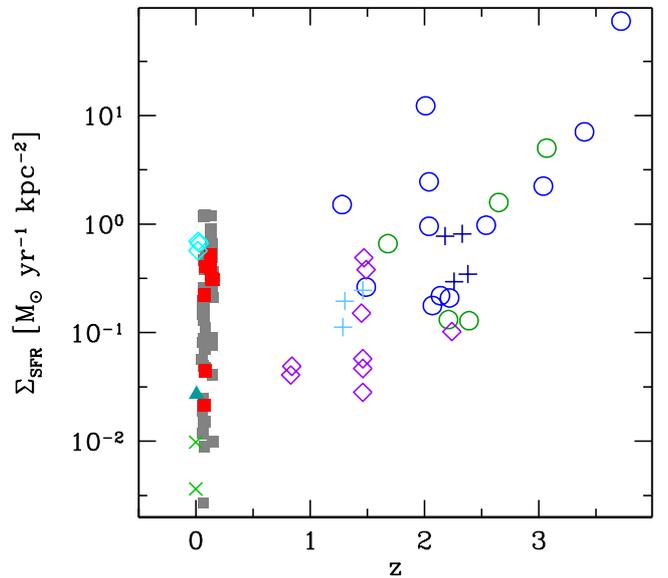}
\end{center}
\caption{ The galaxy star formation rate surface densitty
  ($\Sigma_{SFR}$) is plotted against redshift for the galaxies from
  Fig.~\ref{fig:ms}. The symbols are the same as in Fig.~\ref{fig:ms}.
  The star formation rate surface densities of DYNAMO-{\em HST}
  galaxies are similar to that of clumpy galaxies ranging
  z$\sim1.0-2.2$. The star formation rate surface density for local
  spirals M83 \& M51, as well as the Antennae system are significantly
  lower than both DYNAMO galaxies and high-redshift
  galaxies.  }
  \label{fig:sigz} 
\end{figure}

\section{Sample}

\subsection{DYNAMO Survey}

Targets for the present investigation were chosen from the
DYNAMO sample which is described in detail in \cite{green2014}.

DYNAMO galaxies were selected from the Sloan Digital Sky Survey (SDSS)
DR4 \citep{sdssdr4}. Global physical properties for the DYNAMO sample
were taken from the JHU-Garching Value Added Catalogues
\citep{brinchmann2004,kauffmann2003,tremonti2004}, and the
sample consists of star forming galaxies that do not show signs of
significant active nuclei, based on the BPT diagram. 
Galaxies were chosen to lie within two redshift windows (centered
at $z\sim0.075$ and $z\sim0.13$ ) to avoid
atmospheric absorption at the wavelength of redshifted H$\alpha$.
\begin{figure*}
\begin{center}
\includegraphics[width=0.85\textwidth]{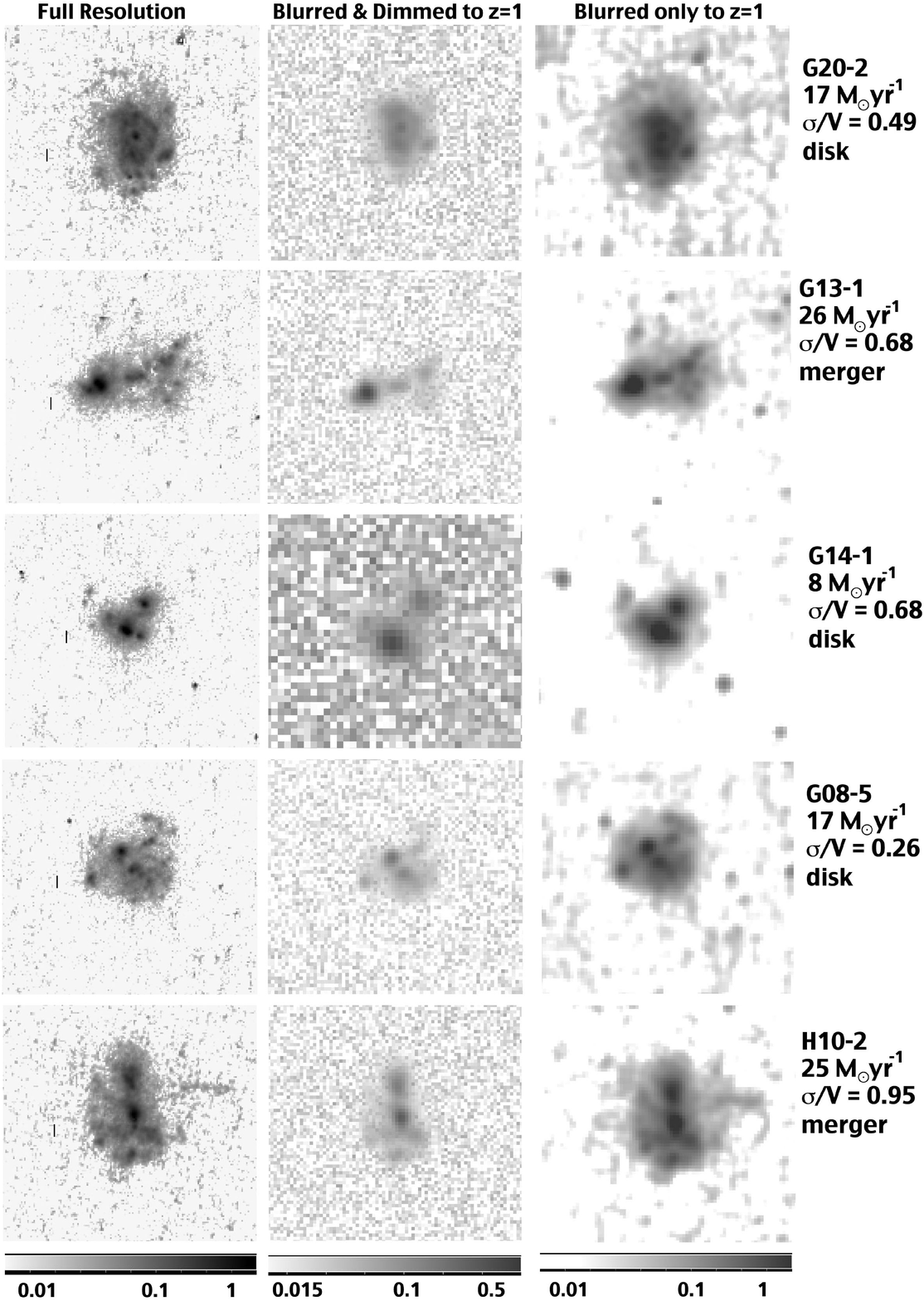}
\end{center}
\caption{The continuum subtracted H$\alpha$+[NII] maps for the DYNAMO
  sample galaxies are shown here. Left column shows full resolution
  maps. The middle column shows maps that have been degraded to
  simulate $z\sim1$ observations with blurring, surface brightness
  dimming and a sensitivity cut similar to high-$z$ AO observations.
  The right column shows only the effect of blurring maps to match
  $z\sim 1$ resolution. The FWHM of the blurring corresponds to
  1.6~kpc and pixel size 0.8~kpc. The maps are arranged according to
  measured gas velocity dispersion ($\sigma_m(H\alpha)$ from
  \protect\citealp{green2013}) from highest (top left) to lowest (bottom
  right). The color bar shows the units of flux in
  $10^{-18}$~erg~s$^{-1}$~\AA$^{-1}$~cm$^{-1}$. Also the galaxy name, total star
  formation rate, $\sigma/V$ and optical morphological classification
  is listed on the far right. A white line is plotted in each panel
  indicating 1~kpc. The DYNAMO sample shows clumpy structures on the
  scale of a few hundred parsecs. We show the same brightness scale
  for all maps to emphasize the stark difference in clump prominence
  between the star forming turbulent disks and the two galaxies with
  properties more like local spirals (eg.~A04-3). Clumps become less
  prominant with decreasing velocity dispersion and decreasing star
  formation rate.  
}
\label{fig:maps}
\end{figure*}
\setcounter{figure}{2}
\begin{figure*}
\begin{center}
\includegraphics[width=0.85\textwidth]{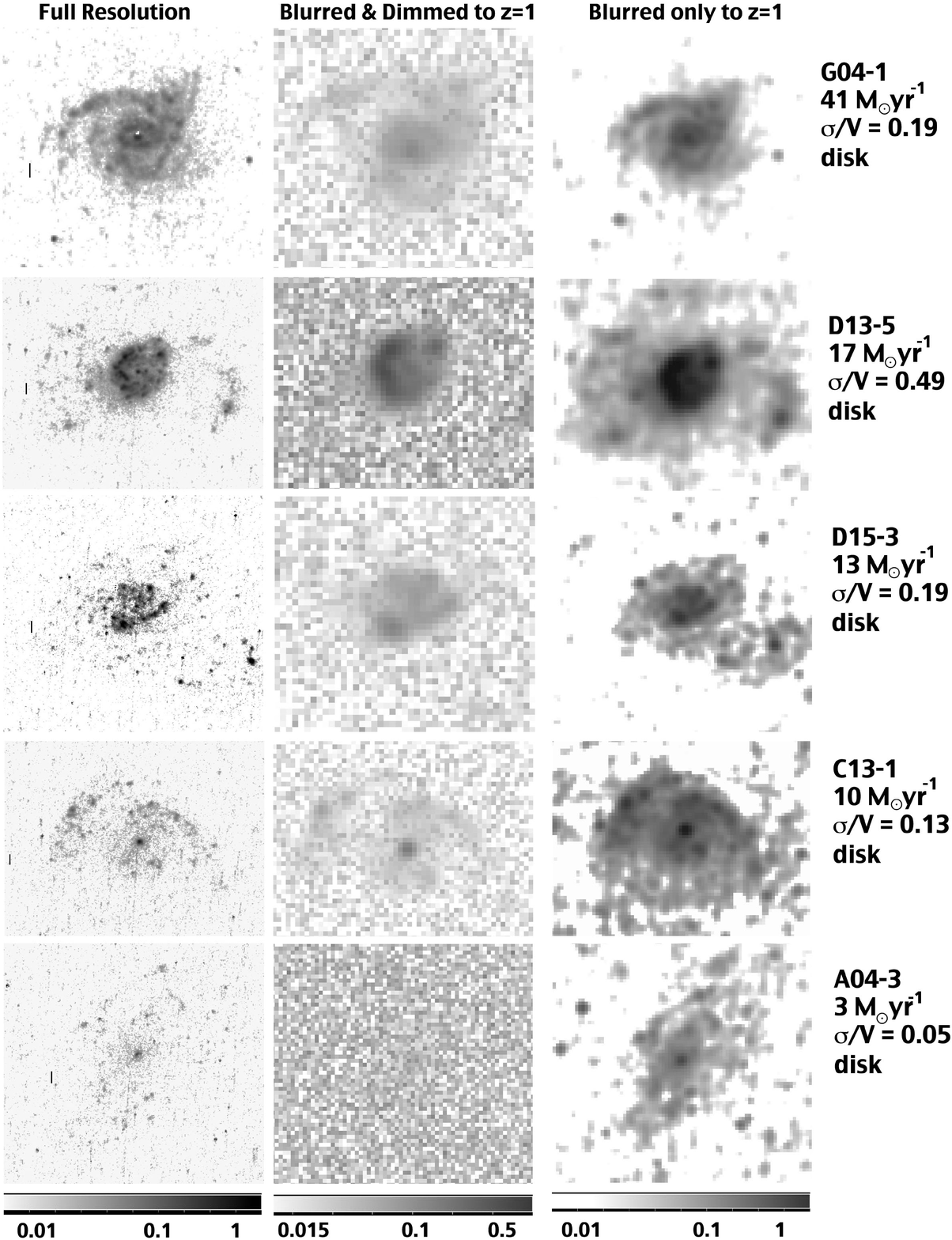}
\end{center}
\caption{The continuum subtracted H$\alpha$+[NII] maps for the DYNAMO
  sample galaxies are shown here. Left column shows full resolution
  maps. The middle column shows maps that have been degraded to
  simulate $z\sim1$ observations with blurring, surface brightness
  dimming and a sensitivity cut similar to high-$z$ AO observations.
  The right column shows only the effect of blurring maps to match
  $z\sim 1$ resolution. The FWHM of the blurring corresponds to
  1.6~kpc and pixel size 0.8~kpc. The maps are arranged according to
  measured gas velocity dispersion ($\sigma_m(H\alpha)$ from
  \protect\citealp{green2013}) from highest (top left) to lowest (bottom
  right). All images are shown in log scale to emphasize the
  substructure. The color bar shows the units of flux in
  $10^{-18}$~erg~s$^{-1}$~\AA$^{-1}$~cm$^{-1}$. Also the galaxy name,
  total star formation rate, $\sigma/V$ and optical morphological
  classification is listed on the far right. A white line is plotted
  in each panel indicating 1~kpc. The DYNAMO sample shows clumpy
  structures on the scale of a few hundred parsecs. Clumps become less
  prominant with decreasing velocity dispersion and decreasing star
  formation rate.   }
  \label{fig:maps}
\end{figure*}

\subsubsection{DYNAMO-{\em HST}}

In this paper we investigate the properties of a sub-sample of DYNAMO
galaxies imaged with the Hubble Space Telescope.  This sub-sample
(hereafter referred to as DYNAMO-{\em HST}) were chosen to be extremely
bright in H$\alpha$ luminosity: $L_{H\alpha}>10^{41.5}$~erg~s$^{-1}$
in an SDSS fiber, with corresponding star formation rates of $\sim$10
-- 40~M$_{\odot}$~yr$^{-1}$. The H$\alpha$ luminosity cut is chose to
match luminosities in of
\cite{forsterschreiber2009,law2009,epinat2009}. All DYNAMO-{\em HST}
galaxies were initially selected to have rotating, disk-like kinematics in
\cite{green2014}, as indicated by significant velocity gradients, and
large gas velocity dispersions in the outer disk,
$\sigma_{H\alpha}\sim 30-80$~km~s$^{-1}$ (measured with integral field
spectroscopy by \citealp{green2013}). Subsequent analysis with higher
resolution data and complimentary surface photometry, after the
HST program was in complete, found that 2 of the targets H10-2 and
G13-1 better resemble mergers (see Appendix for a detailed
discussion).  

The sample spans $M_{star} = 
1-9\times 10^{10}$~M$_{\odot}$, and extinction $A(H\alpha)\sim
0.2-1.7$~mag, both similar to samples of high-$z$ turbulent and clumpy
disks \citep[e.g.][]{forsterschreiber2009,swinbank2012}. Overall,
galaxies similar to those in DYNAMO-{\em HST} are extremely rare in the
local Universe, with a space density of $\sim 10^{-7}$~Mpc$^{-3}$.

In addition to the sample described
above, as a control of our method we also targeted one galaxy
(DYNAMO A04-3) that has properties that are more typical of low-$z$
galaxies. This object has a star formation rate of $\sim 3$~M$_{\odot}$~yr$^{-1}$ and
$\sigma_{H\alpha}\sim 10$~km~s$^{-1}$.

Thirteen galaxies were imaged as part of the DYNAMO-{\em HST} campaign, but
observations of three galaxies did not yield usable data.  In one case
the target was slightly misaligned (spectrally) on the ramp filter, so
significant amounts of emission were not mapped. For two galaxies the
flux was simply too faint (due to their large distance), and we were therefore
unable to identify any substructures with greater than
$\sim 1\sigma$ confidence. The final DYNAMO-{\em HST} sample
therefore consists of the ten galaxies whose properties are summarized
in Table~1.

Figure~\ref{fig:ms} shows global star formation rate versus stellar
mass ($M_{star}$) for the DYNAMO-{\em HST} sample, along with the
corresponding data for other published samples of clumpy galaxies. In
the present paper, star formation rates were calculated from
extinction-corrected H$\alpha$ line luminosity by assuming star
formation rate~[M$_{\odot}$~yr$^{-1}$] = $5.53\times 10^{-42}
L_{H\alpha} $ [~erg~s$^{-1}$] \citep{hao2011}. We calculate the
intrinsic H$\alpha$ extinction using the H$\alpha$ and H$\beta$ line
ratios from SDSS spectrum.  Open circles in Figure~\ref{fig:ms}
correspond to samples of lensed galaxies. The sample of
\cite{livermore2012} is not shown as stellar masses are not available
for their galaxies; however the star formation rate of the
\cite{livermore2012} sample are mostly $\sim
0.1-1$~M$_{\odot}$~yr$^{-1}$, with one galaxy $\sim
10$~M$_{\odot}$~yr$^{-1}$. It is evident from Figure~\ref{fig:ms} that
the stellar masses of galaxies in the DYNAMO-{\em HST} sample are
useful for studies of clump properties of in galaxies with stellar
masses in the range $\sim1-5\times 10^{10}$~M$_{\odot}$. Galaxies in
this mass range dominated the cosmic star formation rate at $z\sim
1-3$ \citep{karim2011}.

The average half-light radius of H$\alpha$ emission in DYNAMO-{\em
  HST} is $R_{1/2}\sim2.5$~kpc this is similar to what is found in
high redshift galaxies \citep[eg~][]{forsterschreiber2009}. In
Figure~\ref{fig:sigz} we compare the star formation rate surface
density of DYNAMO-{\em HST} galaxies to the same sample of
high-redshift galaxies as in Figure~\ref{fig:ms}. For all galaxies we
calculate the star formation rate surface density using the total,
extintion corrected star formation rate and the H$\alpha$ half-light
radius. DYNAMO galaxies have median $\Sigma_{SFR}\simeq
0.3$~M$_{\odot}$~yr$^{-1}$~kpc$^{-2}$, for the high-$z$ clumpy galaxy
sample we study here the median star formation rate surface density is
very similar.  Galaxies in
\cite{swinbank2012,wisnioski2012,genzel2011} all have median star
formation rate surface densities of
0.06-0.5~M$_{\odot}$~yr$^{-1}$~kpc$^{-2}$.  We note that in
Fig.~\ref{fig:sigz} there is a trend that for $z\gtrsim 2.5$ the
typical star formation rate surface density increases. Those samples
which target higher redshift galaxes, \cite{livermore2015} and
\cite{jones2010}, have median star formation rate surface density of
1.2 and 0.7~M$_{\odot}$~yr$^{-1}$~kpc$^{-2}$, respectively.  Indeed
the results of Figure~\ref{fig:ms}~\&~\ref{fig:sigz} suggest that the
appropriate redshift range for comparison of DYNAMO disks to high-$z$
galaxies is $z\simeq1-2.5$.

\section{Methods}

\subsection{New Observations}
H$\alpha$ flux maps of DYNAMO galaxies were obtained using the Wide Field Camera on
the Advanced Camera for Surveys (WFC/ACS) on the Hubble Space
Telescope ({\em HST}; PID 12977, PI Damjanov).  These 
narrow-band imaging observations were undertaken using the HST ramp filters FR716N and FR782N, which target
the H$\alpha$ emission line with 2\% bandwidth. Central wavelengths in
our sample range from 700-754~nm. Data were also taken with the
associated continuum filter FR647M. Integration times were
45~min in the narrow band filter and 15~min with the continuum
filter. All images were reduced using the standard {\em HST} pipeline.

\subsection{Continuum Subtraction \& H$\alpha$ Maps} 

Our analysis of the clump properties of the DYNAMO-{\em HST} sample requires
the use of continuum-subtracted maps of H$\alpha$+[NII].  In order to
subtract the continuum, we first aligned the emission line (FR716N or
FR782N) and continuum (FR647M) maps using the positions of point
sources.  Adjustments were made iteratively until the point sources
were aligned within a few percent of a pixel. We then convolved each
galaxy's SDSS spectrum with the filter throughput curves for the
medium and narrow band filters, and also convolved each SDSS spectrum
with two narrow bandpasses just blue and red of the narrowband filter
bandpass. This was done in order to estimate the continuum as closely
as possible to the H$\alpha$+[NII] emission lines. To avoid
contamination by the [SII] doublet, we constructed these bands as
follows: First the narrow-band filter throughput was divided into two
equal parts so that integrated throughput in each of these is
0.5. Second, these artificial filters were put then on each side of
H$\alpha$+[NII] and convolved with the SDSS spectrum. Finally, the
flux from artificial filters is summed to give total narrow-band flux coming from the
continuum at the position of H$\alpha$.

The flux ratio of the medium band filter region to the continuum
surrounding H$\alpha$+[NII] was determined, and the flux of the FR647M
filter was then scaled by this ratio. We assume a single correction
for each galaxy. For most targets the single SDSS 3 arcsec fiber
covers the extent of the galaxy. However for larger targets (namely
C13-1 and A04-3) we make the assumption that the correction in central
3 arcsec is sufficient to describe the entire galaxies. We remind the
reader that our targets are selected to not include AGNs.  Finally,
the scaled FR647M was subtracted from the narrow-band filter maps to
construct the final map of H$\alpha$+[NII]. The continuum subtracted
H$\alpha$+[NII] maps for DYNAMO-{\em HST} are shown in
Figure~\ref{fig:maps}. Visual inspection of these maps makes it clear
that many individual clumps have very bright H$\alpha$ emission, as do
some nuclei. Comparison of the continuum flux of clumps will be the
subject of a future paper. 
\begin{figure}
\begin{center}
\includegraphics[width=0.49\textwidth]{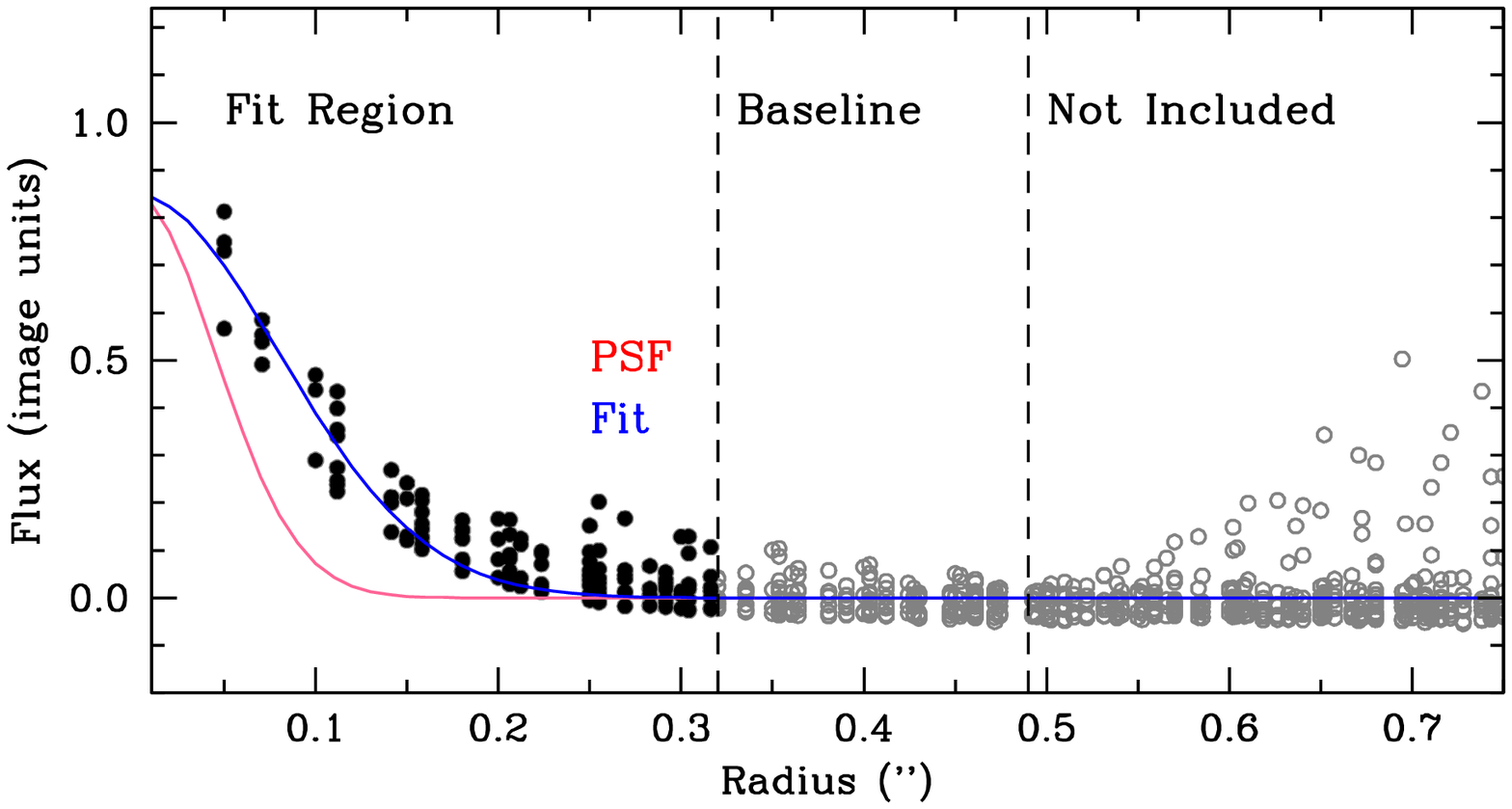}
\end{center}
\caption{ An example light profile of a DYNAMO clumps from galaxy
  D13-5 is shown. The grey points represent those included in the
  Gaussian fit. Radial regions are labeled according to whether they
  are included in the fit, or the baseline determination. The blue
  line represents a 1-D Gaussian fit to the data, and the red line
  represents the beam of the image determined on point sources
  surrounding the galaxy.  }
  \label{fig:clump} 
\end{figure}

\subsection{Identification of clumps} 

The precise definition of what a massive star forming clump is, and
how it should be identified, varies in the literature
\citep[eg.][]{jones2010,genzel2011,livermore2012,guo2015}. In the
present paper we adopt an automated clump identification technique
whose purpose is to isolate clumps in a manner similar to the way they
have been isolated in high-$z$ H$\alpha$ maps of clump galaxies
\citep[eg.][]{genzel2011}.  Our procedure is a variation of the clump
identification strategy employed by \cite{guo2015}, with some minor
changes that find their origin in the fact that our observational
parameters are not identical to theirs (e.g. our resolution is higher,
and our data target H$\alpha$ emission rather than UV emission from
young stars \citealp{guo2015})

The fist step in our procedure was to use an unsharp masking technique
to identify emission peaks that are brighter by a factor of at least
three compared to the emission in smoothed maps of the same galaxy.
Each full-resolution image was divided by a blurred image constructed
by convolving the original image with a 1~arcsec Gaussian beam ($\sim
10\times$ the beam width of {\em HST}).  A simple cut was then used to
identify the locations of all independent peaks that meet our signal
enhancement threshold.  For identification purposes we did not place
any restrictions on the sizes of the clumps, since our goal is to
measure these sizes.  Our methodology passes the simplest sanity check
of having the brightest set of peaks in the H$\alpha$ emission maps
always corresponding to clumps.

In the sample of DYNAMO-{\em HST} galaxies we detected 113 H$\alpha$ peaks
whose fluxes were greater than 3$\times$ that of the surrounding area.
This corresponds to roughly 11 clumps per galaxy, which is slightly
higher than the number of clumps per galaxy reported by
\cite{livermore2012}, who found $\sim 7$ clumps in each of their
lensed galaxies with similar resolution in the direction of maximum
magnification to that of our measurements.  For comparison we blurred
an archival HST/WFC map of H$\alpha$+[NII] emission from the nearby
galaxy M51 to match the physical resolution of a DYNAMO-{\em HST}
observation. We then carried out the same clump identification
procedure and found 67 emission
peaks. 

\subsection{Measurement of Clump properties}

To estimate the sizes of star forming clumps, we fitted elliptical Gaussian
functions to the 2-D brightness distributions surrounding each peak in
the H$\alpha$+[NII] map. This method has been use to estimate the
sizes of H{\sc ii} regions for many years (e.g. \cite{kennicutt1979}).  The
reader is referred to
\cite{wisnioski2012} for a detailed discussion of how this method 
can best be used to estimate
the sizes of massive star forming clumps, and to a comparison of
the sizes estimated using this technique to sizes determined using isophotal radii.

Figure~\ref{fig:clump} shows an example radial light profile of a
clump from DYNAMO galaxy D13-5. Before fitting the Gaussian, we subtracted
the surrounding emission, and this baseline flux was determined
iteratively. The initial starting point for this procedure was
determined by visual estimation, and this background value was then subtracted from 
the entire galaxy. For each clump measurement, flux from
surrounding clumps was masked, and an initial 2-D Gaussian fit was produced.
The baseline background was then
re-determined using a region $R> 4\times r_{clump}$, where
$r_{clump}$ is the half-width of the major axis of the fitted
ellipse. The fit was then recomputed iteratively until the fit parameters
converged.

The radius of each clump was defined to be the geometric mean of the
standard deviations of the major and minor axes ($a$ \& $b$
respectively) of the 2-D Gaussian function, {\em i.e.} $d_{core} =
2\times\sqrt{a\times
  b}$. For most clumps this diameter is almost identical to the value
returned from a 1-D Gaussian fit to the brightness profile of the
clump. However, in cases of very non-circular clumps the size definition chosen
always contains the same fraction of the flux, independently from 
the clump ellipticity. The resolution of each map was measured using
ten point sources in the map surrounding each galaxy. We chose to
measure this empirically to account for any blurring that was induced
by the continuum subtraction. The median beam size of our maps is
0.088 arcsec, which is 160~pc at $z=0.1$. The final clump sizes quoted in this paper have
had this beam size subtracted in quadrature from the clump sizes
determined from the H$\alpha$+[NII]
maps.

Clump fluxes were calculated by integrating the light in the
H$\alpha$+[NII] map through an elliptical aperture that is $3\times$
the size of the core diameter of the fitted ellipse. In the event that
a nearby clump overlapped with this region, we masked the nearby clump
to calculate the clump properties. For the flux calculation of clumps
with masked regions we then interpolate the flux in the missing region
using the radial profile. This affects the clump flux at the 10-20\%
level.  Measured fluxes were then adjusted for extinction, [NII]
contribution (determined from SDSS spectra), and distance. The
redshift, extinction correction and [NII] correction are each given in
Table~\ref{table:sample}. As mentioned previously, star formation
rates were calculated from extinction-corrected H$\alpha$ line
luminosity by assuming SFR~[M$_{\odot}$~yr$^{-1}$] = $5.53\times
10^{-42} L_{H\alpha} $ [~erg~s$^{-1}$] \citep{hao2011}.

Finally, because the properties of central clumps in a galaxy are
likely affected by different physical mechanisms (eg. gas in flow to
center of galaxy potential), for the analyses in the present paper we
have omitted clumps that were coincident with the galaxy centers
determined from the starlight (FR647M) maps. Properties of individual
clumps in  DYNAMO galaxies are given in Table~2.

\subsection{Degrading DYNAMO maps to simulate $z\sim 1-2$ observations}

To create a straightforward comparison of the sub-galactic properties
of star forming regions we convolve the DYNAMO maps with a Gaussian
whose width is set to match the physical resolution of AO enabled IFU
observations on $z\sim 2$ galaxies.  Observations of rest frame
H$\alpha$ using AO enabled instruments typically obtain
$\sim$0.1-0.3~arcsec resolution
\citep[eg.~][]{genzel2011,wisnioski2012}. Across the redshift range
$z\sim 1-2$ a median $FWHM \sim 0.2$~arcsec corresponds to $\sim$2~kpc
(or a Gaussian standard deviation of $\sim$800~pc). Note that due to
cosmological effects the difference in spatial resolution from z=1.0
to $z$=2.5 is only $5\%$.  The typical resolution of our DYNAMO-{\em HST}
H$\alpha$+[NII] maps is much smaller than this. In the subsequent
analysis we will use these blurred DYNAMO maps for comparison to
high-$z$ AO samples. We will also use these blurred maps to study how
degrading resolution affects measured clump properties.

We also simulate the effects due to decreased sensitivity of high
redhsift observations. The includes both the $(1+z)^4$ surface
brightness dimming and the decreased sensitivity of AO observations of
high-$z$ H$\alpha$ observations. We apply sensitivity cut that is
chosen to match the median sensitivity of the H$\alpha$ maps in SINS
survey, $5\times 10^{39}$~erg~s$^{-1}$~kpc$^{-2}$. We fit clump
properties both with and with-out the surface brightness effects, and
we find that clumps in maps with the $(1+z)^4$ cosmological surface
brightness dimming to $z=1$ and sensitivity cut have measured core diameters
that are the same within 10\% of the blurred core
diameter. \cite{calvi2014} find a similar result. Because clumps are
such high surface brightness features they are less effected by
redshift dimming. The largest impact is in the ability to detect the
clump, we will discuss this in detail later. 

Throughout the paper, if a clump is detected in the blurred and dimmed
maps then we will use the values from that map. If the clump is only
detected in the ``blurred only'' maps and not in the dimmed map, we
will use the blurred only value. This is justified since we find that
surface brightness effects have on average no effect on clump sizes,
and thus by including these we can study blurring effects with better
statistics. Clumps detected in ``blurred only'' and ``blurred +
dimmed'' maps will be denoted with seperate symbols.

Note that clumps are identified independently each in the blurred
DYNAMO maps (Fig.~\ref{fig:maps} right column) and the full resolution
maps (Fig.~\ref{fig:maps} left column). We do not use the same clump
identifications for each. This is done so that we can study, later in
this paper, the effect of resolution on clump identification.
\begin{figure}
\begin{center}
\includegraphics[width=0.5\textwidth]{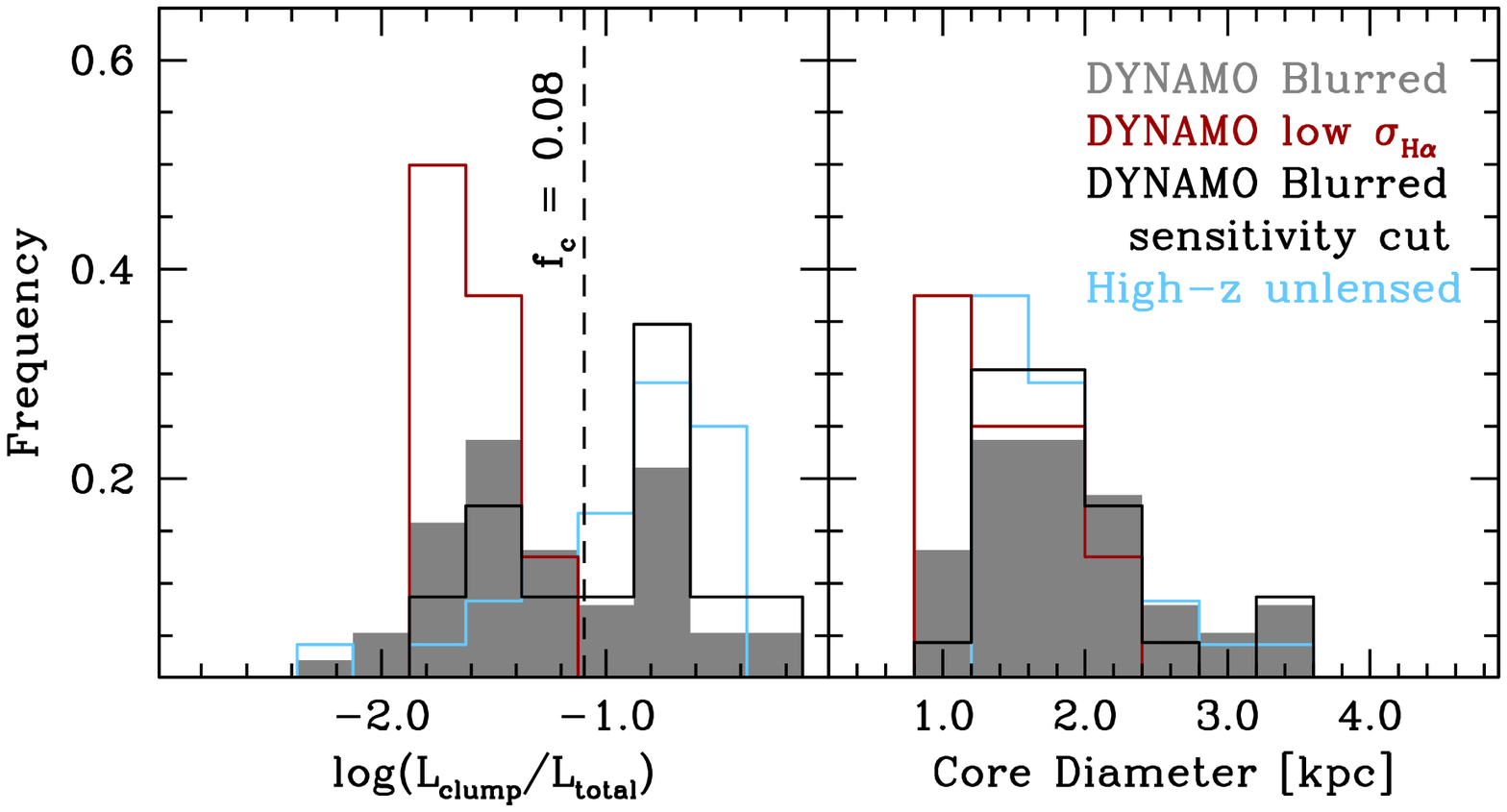}
\end{center}
\caption{Distribution of the fractional H$\alpha$ luminosity for each clump to
  that of the total galaxy, and the distribution of the size of clumps measured in the
  blurred DYNAMO maps, and the high-$z$ unlensed comparison sample
  (light blue histogram). The DYNAMO control galaxy (A04-3, red histogram) is counted
  seperately from the main DYNAMO sample  (grey  shaded region). The
  black line represents the DYNAMO clumps in which a sensitivity cut
  of $5\times10^{39}$~erg~s$^{-1}$~kpc$^{-2}$ has been applied to
 mimic that of the SINS survey \protect\citep{forsterschreiber2009}.  The
  vertical line in the panel showing $L_{clump}/L_{Total}$  marks the
  8\% line. In both cases the
  properties of the DYNAMO clumps are well matched to the properties
  of clumps in high-$z$ galaxies, especially when the sensitivity cut
  is applied to the DYNAMO sample. 
}
\label{fig:z2props}
\end{figure}
\begin{figure*}
\begin{center}
\includegraphics[width=0.8\textwidth]{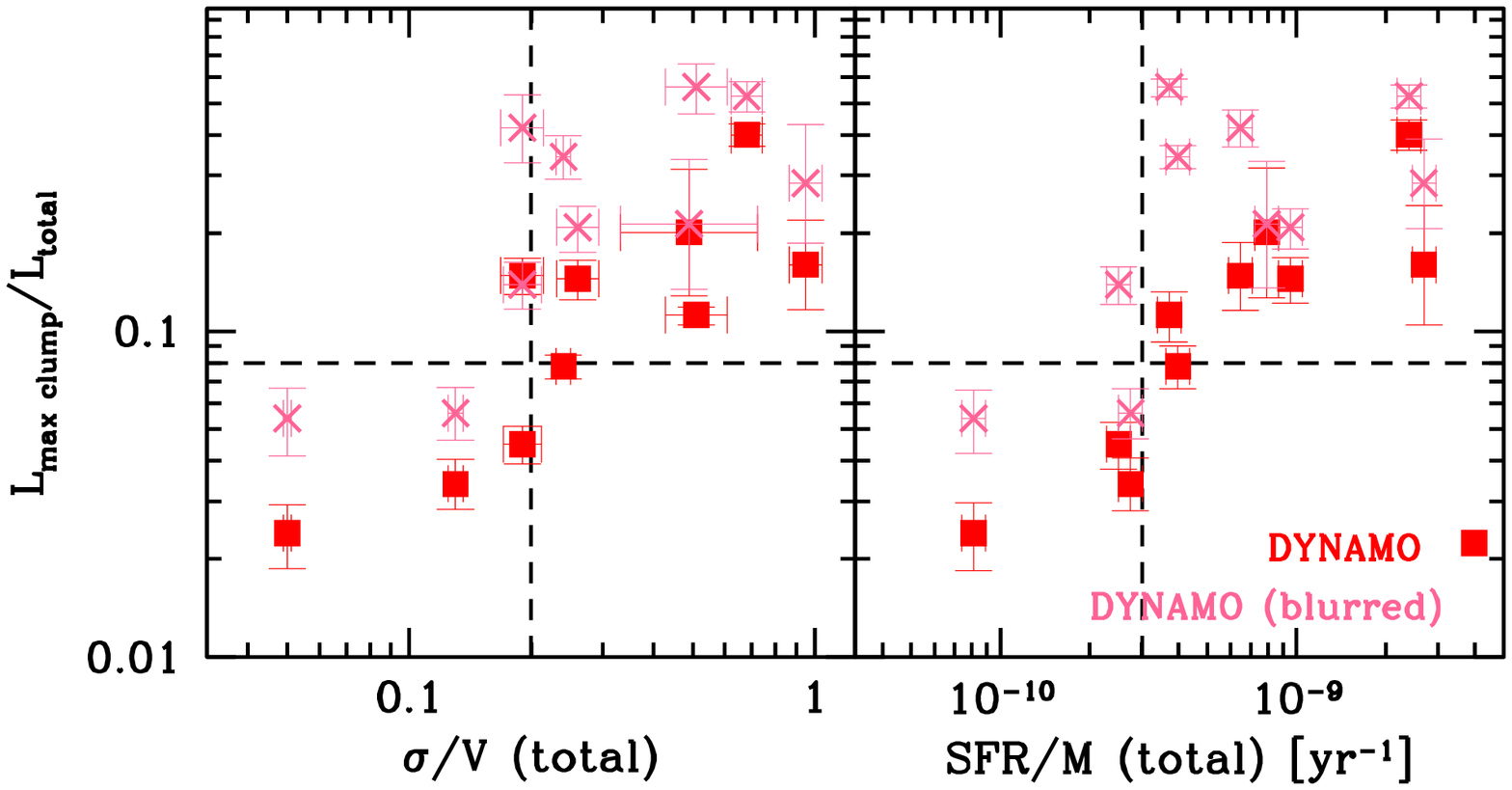}
\end{center}
\caption{ The clump-to-total luminosity ratio of the brightest
  clump in each galaxy is plotted against both the ratio of $\sigma/V$
  and the specific star formation rate for DYNAMO galaxies. The solid
  red squares represent clumps from the full resolution DYNAMO maps,
  and the pink crosses represent the same measurement for the
  DYNAMO maps that have been blurred to match $z\sim 2$
  resolution. The horizontal line represents 8\% of the total
  luminosity. The vertical lines are set to roughly match the point at
  which the high resolution DYNAMO maps intersect with 8\% line. There is a positive correlation between both the
 ratio of velocity dispersion to rotation velocity and the specific star formation
  rate. 
}
\label{fig:clumpiness}
\end{figure*}

\section{Comparison to high redshift main-sequence
  galaxies and local analogues}

\subsection{Comparing DYNAMO galaxies to common definitions of
  ``clumpy'' galaxies}

The DYNAMO clumps have H$\alpha$ luminosities as bright as $\sim
10^{42}$~erg~s$^{-1}$, corresponding to $\sim
5-10$~M$_{\odot}$~yr$^{-1}$. The median DYNAMO clump has
$L_{clump}(H\alpha) \sim 7.5\times10^{40}$~erg~s$^{-1}$, or
$\sim$0.5~M$_{\odot}$~yr$^{-1}$.  To put this in context, the entire
H$\alpha$ luminosity of M~51 ($\sim 2\times 10^{41}$erg~s$^{-1}$), a
nearby star-forming disk, may be fainter than a single, bright H$\alpha$ clump
in a DYNAMO galaxy. In the full resolution maps of the DYNAMO-{\em
  HST} galaxies the median clump size for all clumps is $d_{core} \sim
330$~pc. Only one clump in the entire sample measures a
$d_{core}<1$~kpc.

\cite{guo2015} show, using the CANDELS sample, that defining clumpy galaxies as those in which at
least one off-center clump is greater than 8\% of the light is able to
distinguish high-$z$ star forming main sequence galaxies from nearby
spirals. An important caveat is that this definition is based on UV
light from young stars; in this work we have observed DYNAMO galaxies in H$\alpha$
emission from star forming regions. In the DYNAMO-{\em HST} sample 8 out of 10 galaxies satisfy the CANDELS
definition of clumpy galaxies. The 2 galaxies that do not satisfy the
``clumpy'' definition are C~13-1 and A~04-3; these galaxies have the
lowest gas velocity dispersions of all galaxies in our sample
($\sigma_{H\alpha} \sim$29 \& 10~km~s$^{-1}$ respectively). They also
both have the lowest specific star formation rates in the sample, and therefore are
likely to have the lowest gas fractions. Consistent with a basic
premise that clumpy star formation is driven by turbulent gas with
high gas fractions, those DYNAMO disk 
galaxies with large gas velocities dispersions ($\sigma_{H\alpha} >
40$~km~s$^{-1}$) and high specific star formation rates (SFR/M$_{*}>
3\times 10^{10}$~yr$^{-1}$) are found to be ``clumpy'' galaxies.

In Fig.~\ref{fig:maps} we compare the properties of clumps in the
blurred DYNAMO maps to those published in $z\sim 1-3$ galaxies. In
Fig.~\ref{fig:z2props} the
'high-$z$ unlensed' sample refers to the three studies of clump properties in
non-lensed galaxies discussed above \citep{genzel2011,wisnioski2012,swinbank2012}. 

We identify a number of clumps that are comparable in
$L_{clump}/L_{total}$ to those observed in the high-$z$ Universe.
Indeed the distribution of normalized clump fluxes in DYNAMO galaxies,
after applying the sensitivity cut, is very similar those observed in
high-$z$ galaxies.  In Fig.~\ref{fig:z2props} we show that the
distribution of sizes of DYNAMO clumps (grey shaded region) in blurred
images closely matches that of the high-$z$ unlensed galaxies. We will return
to the effects of resolution on clump sizes.

The black lines in
Fig.~\ref{fig:z2props} represent the distributions of clump properties
that derive from maps in which the sensitivity cut is applied. The
average flux ratio ($<L_{clump}/L_{total}>$) for clumps in the
high-$z$ unlensed sample is $16\%$. The average flux ratio for the
sample of DYNAMO-{\em HST} clumps that have been blurred and pass the SINS
sensitivity cut is almost exactly the same, 17\%. A significant
minority (6/24) of clumps in the high-$z$ unlensed sample have
$L_{clump}/L_{total} > 0.25$, this is similar to the clumps in the
blurred DYNAMO-{\em HST} sample that have passed the sensitivity cut
(4/22). After this
sensitivity cut the 2 DYNAMO galaxies, C13-1 and A04-3 do not have
observable clumps.

In the right panel of Fig.~\ref{fig:z2props} we show the distribution
of clump sizes in the blurred DYNAMO-{\em HST} maps and also the clump sizes in
high-$z$ unlensed galaxies. Blurred DYNAMO clumps clearly span a
similar range and have similar distribution as clumps in high-$z$
galaxies. 

In Fig.~\ref{fig:clumpiness} we show how the clumpiness of DYNAMO
galaxies is related to both the ratio of H$\alpha$ velocity dispersion
to rotation velocity ($\sigma/V$) and the specific star formation
rate. Here we trace ``clumpiness'' with the clump-to-total luminosity
ratio of the brightest clump in each galaxy, to preserve the concept
in the CANDELS definition of a ``clumpy'' galaxy. The pink x's mark
the values for the blurred DYNAMO clumps, and are thus the most
appropriate to compare with high redshift studies. In We find a
positive correlation between the clump-to-total flux ratio of the
brightest clump and both the specific star formation rate
(SFR/M$_{*}$) and $\sigma/V$. \cite{shibuya2015arxiv} show that the
fraction of clumpy galaxies increases with increasing specific star
formation rate. Their result is very similar to the right panel of
Fig.~\ref{fig:clumpiness}, showing again that DYNAMO galaxies are
behaving similarly to $z\sim 1-2$ main-sequence galaxies.
\begin{figure*}
\begin{center}
\includegraphics[width=0.99\textwidth]{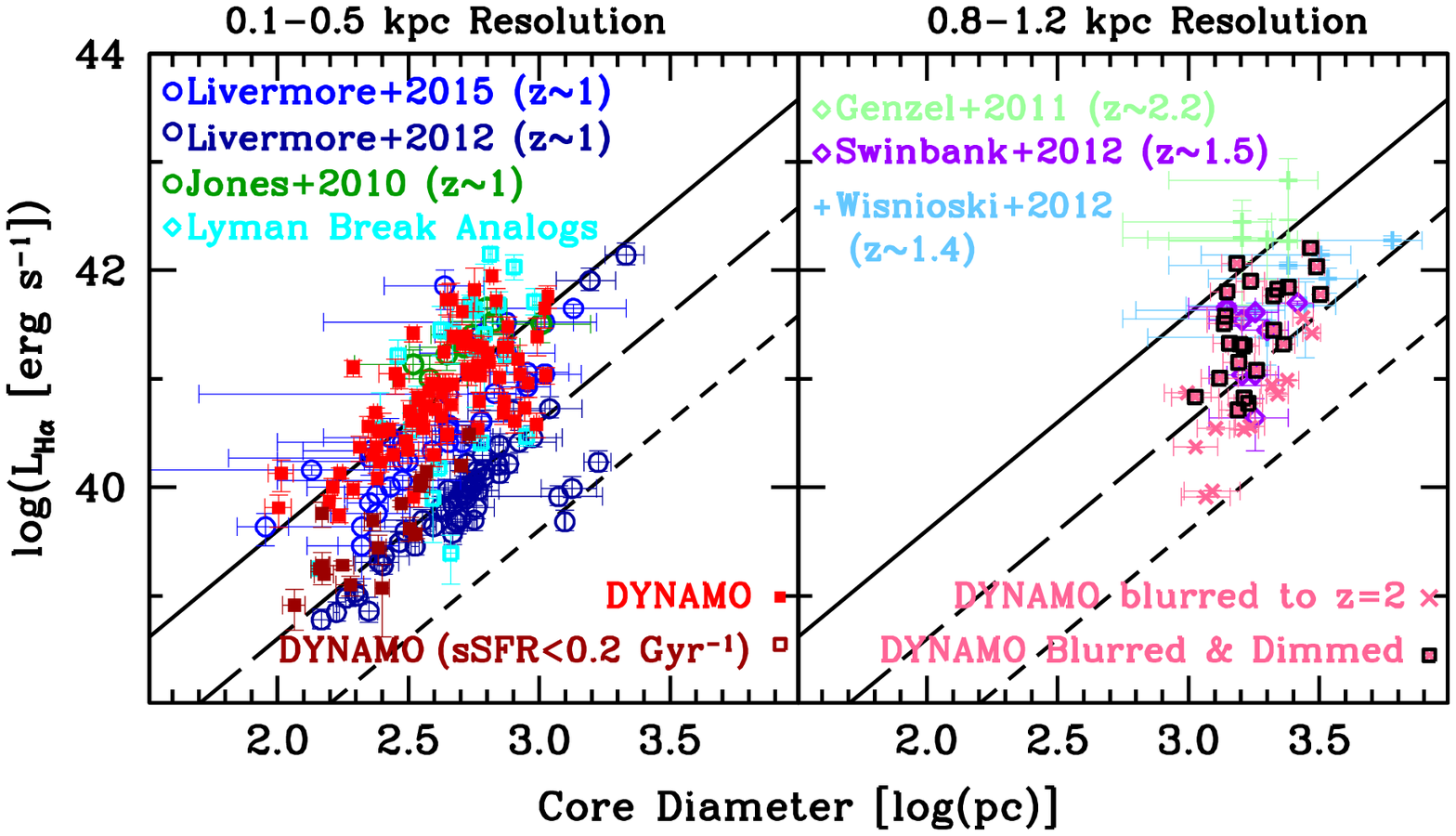}
\end{center}
\caption{ Comparison of position in size-luminosity diagram of DYNAMO
  clumps to clumps in  high redshift galaxies. The {\bf left panel} shows
  those samples in which the resolution is of order $\sim 100$~pc,
  including lensed galaxies, full resolution DYNAMO-{\em HST} galaxies, and
  a sample of nearby Lyman Break Analogues.  The diagonal lines
  indicate lines of constant surface brightness that are set to match
  the median DYNAMO clump (solid line) and then a factor of 1/10 (long
  dashes) and 1/100 (short dashes) lower surface brightness than the
  DYNAMO median. The {\bf right panel} shows those samples of high-$z$
  galaxies in which the spatial resolution is of order $\sim 1
  kpc$. In this panel we also show the clump properties measured on
  the blurred H$\alpha$ maps of DYNAMO, with resolution of $\sim
  800~pc$, as they would if observed with {\em HST} at $z\sim 2$. 
 Symbol colors and shapes are as described in the
  key. Clumps in DYNAMO galaxies are as bright as many of the clumps
  in the unlensed samples of high-$z$ galaxies. DYNAMO clumps are
  brighter than many of the clumps in lensed galaxies, however are
  similar sizes. 
}
\label{fig:radlumhi}
\vskip 24pt
\end{figure*}

In the following sections when we use the term``clumpy'' galaxies from
the DYNAMO-{\em HST} sample, we will mean those that have clumps that
would be detected in the degraded maps, with the median sensitivity
level in the SINS survey, and also meet the criteria described in
\cite{guo2015} for the CANDELS survey. To be explicit those DYNAMO
galaxies are D13-5, D15-3, G04-1, G08-5, G13-1, G14-1, G20-2, and
H10-2. Those galaxies that do not satisfy this definition are A04-3
and C13-1.

\subsection{The size-luminosity correlation of H$\alpha$ clumps}

For those DYNAMO galaxies that satisfy the definition of ``clumpy'',
the median core diameter of clumps when measured at the full
resolution is $\sim$450~pc. The majority (67\%) of clumps have core
diameters that range from 200-650~pc in size. Fewer than 10\% of
clumps in our sample have clumps that are larger than
$d_{core}>800$~pc, and only 2 out of 79 clumps in clumpy DYNAMO
galaxies are larger than 1~kpc.

In Fig.~\ref{fig:radlumhi} we compare the properties of DYNAMO clumps
to those in $z\sim 1.0-2.5$ galaxies.  The size of DYNAMO clumps, when
viewed at full resolution, are very similar to those measured in lensed
galaxies, which have similar resolution due to magnification, $\sim
100-200$~pc. DYNAMO clumps tend to be brighter, than those clumps in
the \cite{livermore2012,livermore2015} samples. Therefore DYNAMO
clumps, on average, have higher star formation rate surface density than the clumps
typically observed in lensed galaxies. 

The star formation rate surface density of clumps in DYNAMO galaxies
is mostly contained within $\Sigma_{SFR}\sim
1-10$~M$_{\odot}$~yr$^{-1}$~kpc$^{-2}$.  The median star formation
rate surface density of DYNAMO clumps (excluding star forming regions
in C13-1 and A04-3) is $\sim 2.3$~M$_{\odot}$~yr$^{-1}$~kpc$^{-2}$. If
we combine the samples of lensed galaxies from \cite{jones2010},
\cite{livermore2012} and \cite{livermore2015} the median star formation rate surface density  of star forming regions is $\Sigma_{SFR}(lens)\sim
0.5$~M$_{\odot}$~yr$^{-1}$~kpc$^{-2}$, close to an order of magnitude
lower than in clumpy DYNAMO galaxies. This is largely due to the
samples of \cite{livermore2012,livermore2015}, which target galaxies
with lower total star formation rate. The median star formation rate surface density of clumps in the
\cite{jones2010} sample is $\sim 2$~M$_{\odot}$~yr$^{-1}$~kpc$^{-2}$,
similar to the DYNAMO sample.

In summary, Fig.~\ref{fig:radlumhi} shows that the clumps in DYNAMO
galaxies are consistent in size and luminosity with those of
high-redshift galaxies, as well as clumps other galaxies that are considered
analogues of high redshift galaxies (LBAs). 
\begin{figure}[t]
\begin{center}
\includegraphics[width=0.45\textwidth]{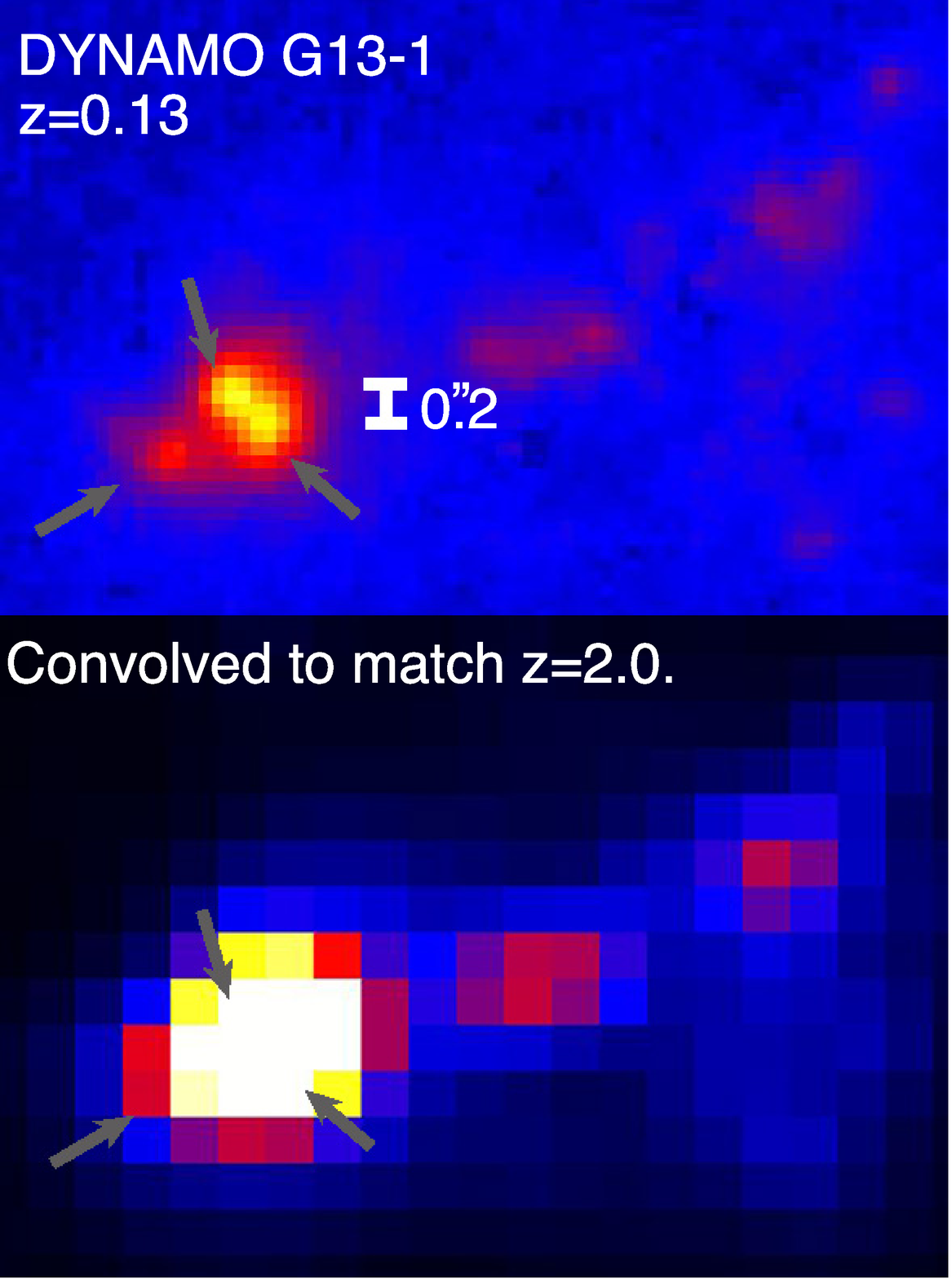}
\end{center}
\caption{ Top: DYNAMO galaxy G13-1 at the full
  resolution of {\em HST} for a z=0.13 galaxy. Bottom: The same galaxy, but the map has been blurred to simulate the typical
  resolution on AO enhanced (or {\em HST}) maps of $z=2$ galaxies (FWHM$\sim 1.8$~kpc). The white bar represents 500~pc. The arrows
  indicate the location of 3 clumps identified in the full resolution
  H$\alpha$+[NII] map, these clumps are blurred to appear as one giant
  clump at $z\sim 2$ resolution. 
  }
\label{fig:g13_1}
\end{figure}
\begin{figure}
\begin{center}
\includegraphics[width=0.48\textwidth]{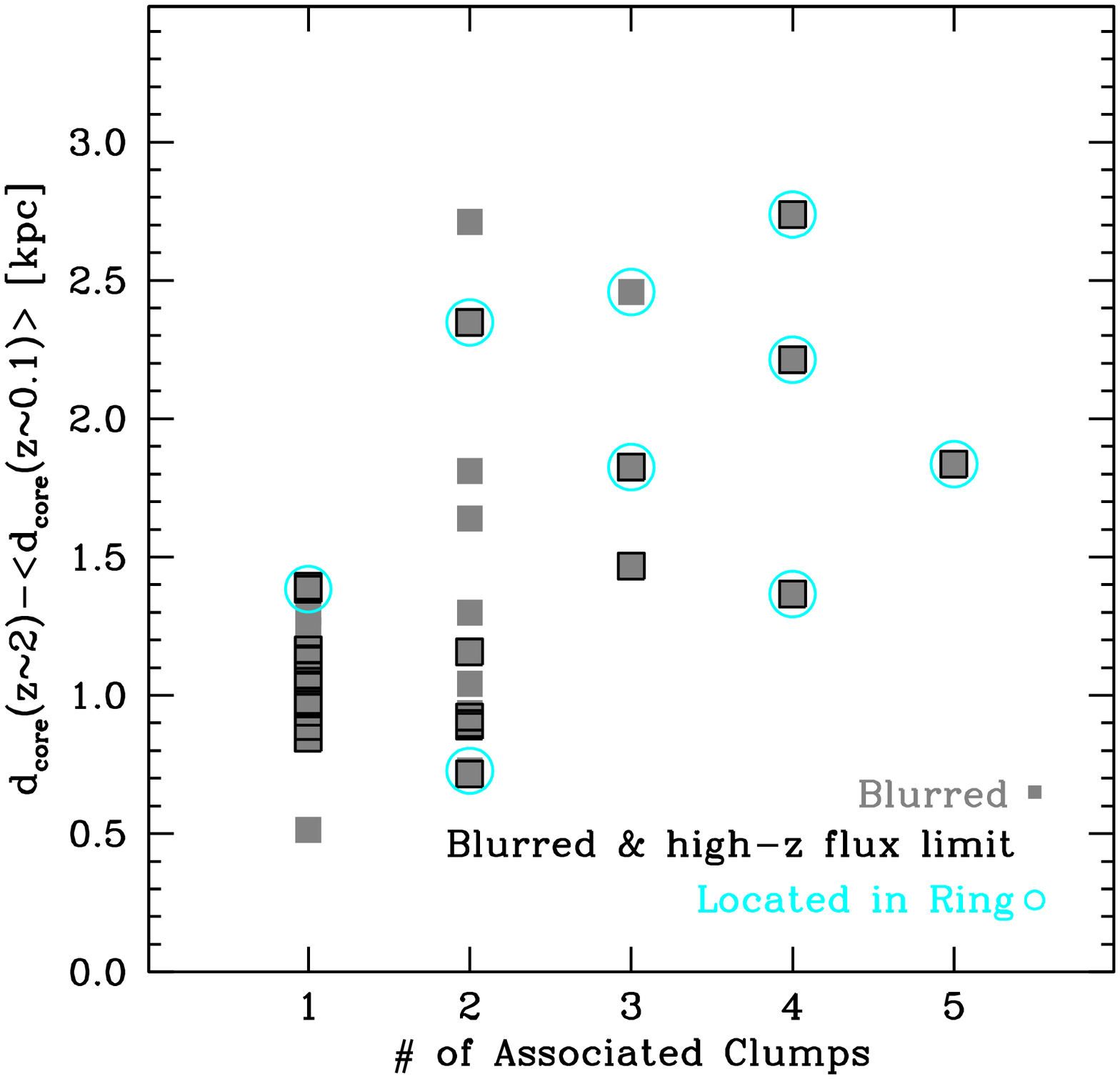}
\end{center}
\caption{ Here we compare the difference in core sizes from the
  blurred to the unblurred maps to the number of clumps in the full
  resolution map that are associated with a clump in the blurred
  maps. Grey squares indicate clumps identified in blurred H$\alpha$
  maps. Black open squares indicate those clumps that
would be observed with the SINS sensitivity cut. Cyan rings indicate
thows clumps that reside in the ring region of the disk. Clumps which
correspond to a single clump on average reflect the difference in the
resolution between the two measurements. However, when clumps in the
blurred maps
correspond to more clumps they are systematically larger. }
\label{fig:deltaR}
\end{figure}
\begin{figure*}
\begin{center}
\includegraphics[width=0.48\textwidth]{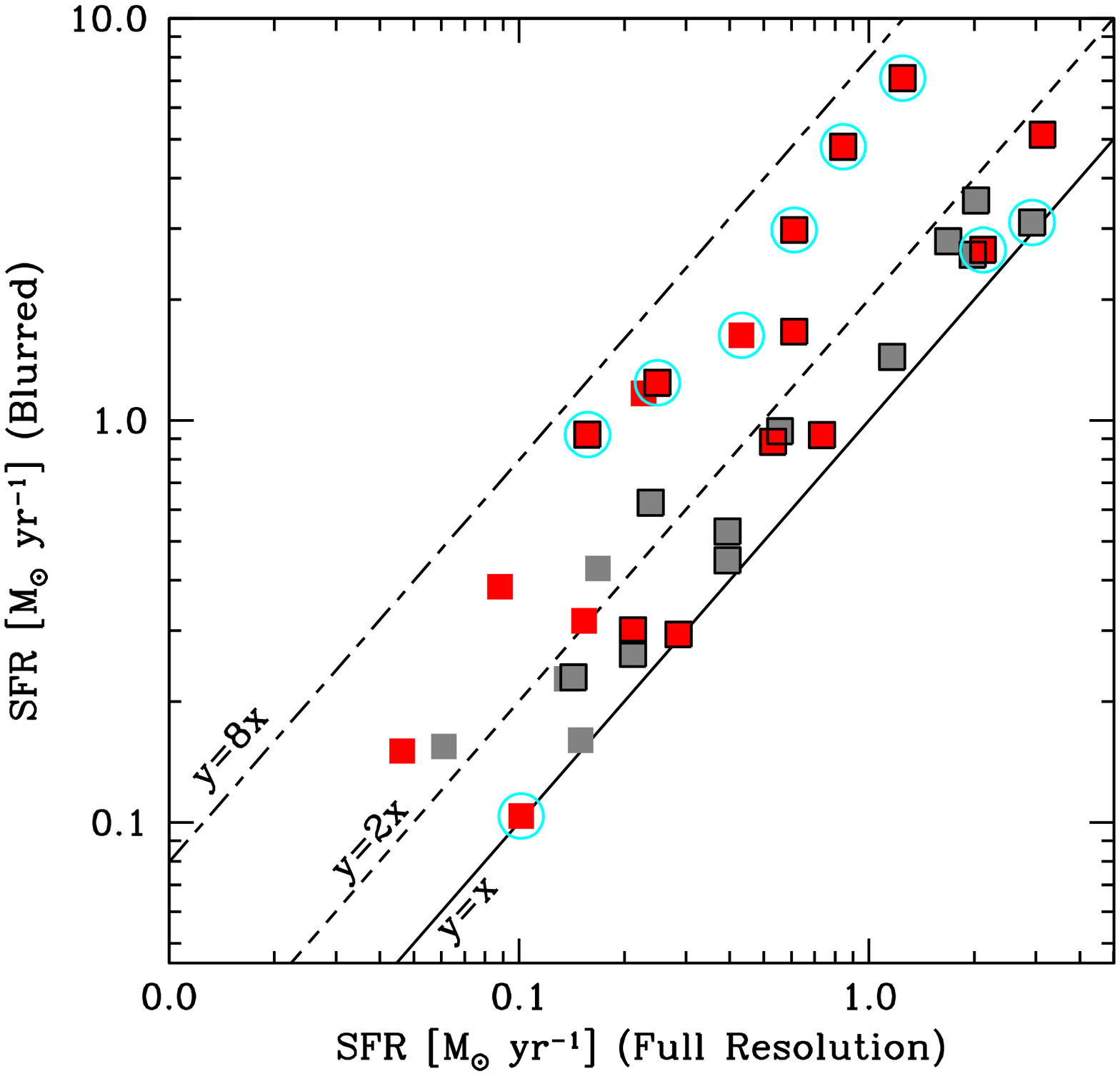}
\includegraphics[width=0.48\textwidth]{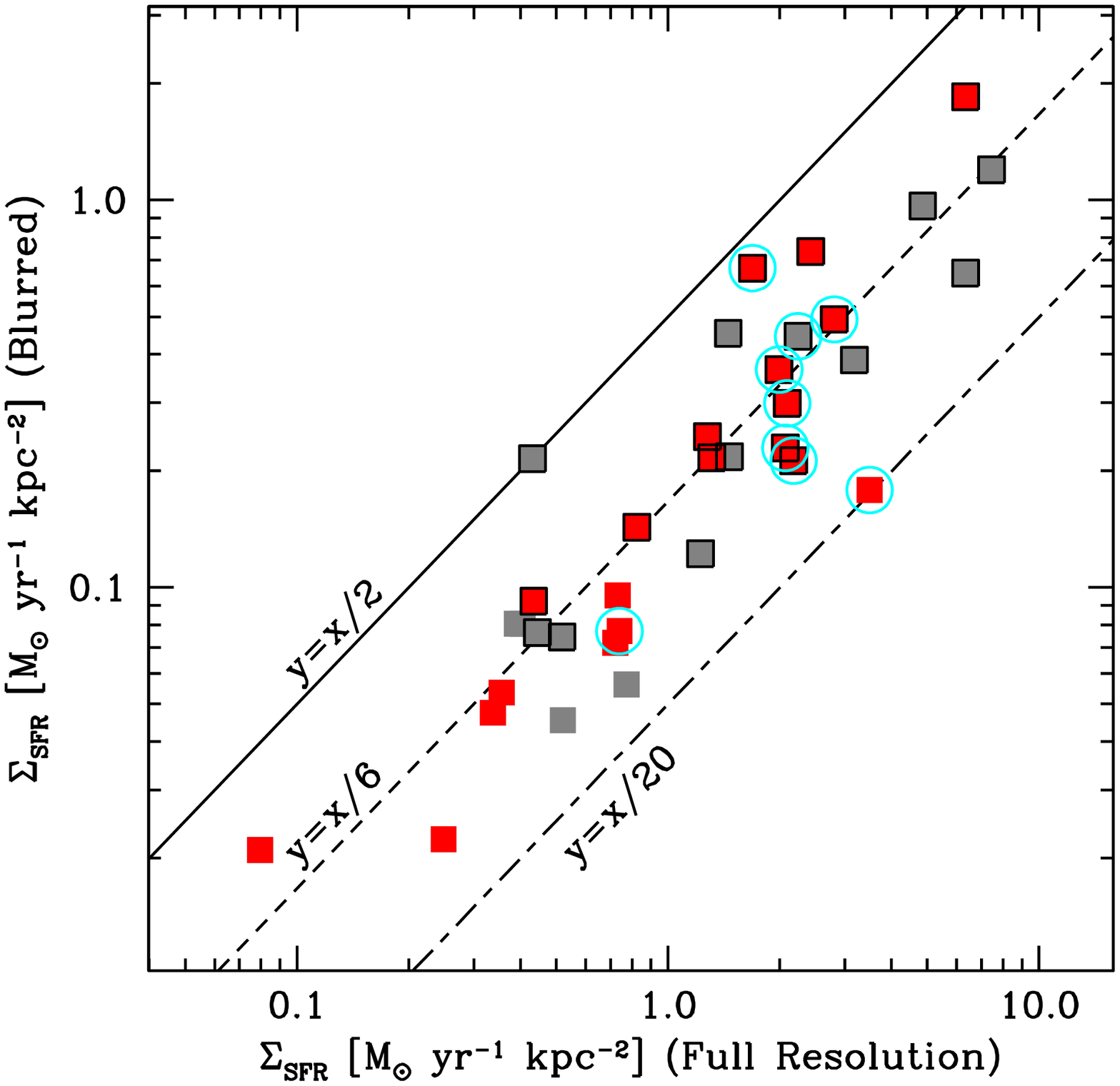}
\end{center}
\caption{  Here we show how degrading images to match $z\sim 2$
  conditions affects the physical properties of individual clumps. The grey squares indicate
blurred clumps which correspond to a single high resolution clump. Red
squares indicate blurred clumps that correspond to more than one clump
in the high resolution map (as illustrated in
Fig.~\ref{fig:g13_1}). 
 The lines indicate scale factor
differences between the two quantities. In the left panel we show the
impact of clump clustering on the derived SFR of blurred clumps in the DYNAMO
sample. When blurred clumps correspond to
more than one clump in the full resolution map and/or when blurred
clumps reside in rings the impact of lower resolution on luminosity is more
severe. In the right panel we show how blurring of clumps affects the
  measured SFR surface density of clumps. The SFR surface density of blurred
  clumps ranges from 1/2 to 1/20 the surface brightness of clumps at
  full resolution. }
  \label{fig:blur_effects}
\end{figure*}

\section{Systematic effect of clump clustering on blurred DYNAMO H$\alpha$ clumps}

In Fig.~\ref{fig:g13_1} we illustrate how the effect of degrading
resolution is not simply to blur the properties of a single clump. In
this figure arrows indicate a complex of 3 clumps, which when viewed
at $z\sim 2$ resolution appears as one clump. At $z\sim0.1$
resolution, the three clumps have an average core diameter of 600~pc,
each being separated by $\sim 400$~pc, when blurred together the
complex has a core diameter of $\sim$2~kpc. This behavior is common
in the maps, and also has the effect of the blurred DYNAMO maps having
significantly fewer clumps.

Both blurring and reduced sensitivity decrease the number of
clumps. In all of the clumpy DYNAMO galaxies we detect 90 clumps at
full resolution (including central clumps and excluding A~04-3 \&
C~13-1), roughly 11 clumps per galaxy. In the blurred maps of the same
set of galaxies we detect only 35 clumps, fewer than 5 clumps per
galaxy. Resolution effects alone account for a decrease in the number
of clumps by a factor of $\sim
2-3$. 
When we account for the $(1+z)^4$ cosmological dimming
and reduced sensitivity for restframe H$\alpha$ surveys of $z\sim1-3$
galaxies that use near-IR spectroscopy the affect on clump numbers
decreases to $\sim 1-4$ per galaxy. 
This smaller rate of clumps in the blurred maps with
a high-$z$ mimicking sensitivity degredation is comparable to the amount of
clumps detected in high-$z$ surveys of unlensed galaxies
\citep{genzel2011,wisnioski2012,swinbank2012}.

When a blurred clump is associated with multiple clumps in the full
resolution map we use the average of the associated clump properties
for comparison. There is very little difference if we choose the
maximum clump size or the average clump size.  

In Fig.~\ref{fig:deltaR} we compare the difference in $d_{core}$ from
blurred-to-full resolution to the number of full resolution clumps
corresponding to each blurred clump. When blurred clumps correspond to
a single clump the difference reflects, on average, the difference
between the resolution elements. As the number of corresponding full
resolution clumps increases as does the affect on the measured size of
the blurred clump. Blurred clumps with very large sizes ($d_{core}
\gtrsim 2$~kpc) are associated with multiple clumps at full resolution
in all cases.

The star formation rate of individual blurred clumps are on average higher than
associated full resolution clumps by 2-3$\times$
(Fig.~\ref{fig:blur_effects}), and in some cases the star formation rate can be
increased as much as $\sim 6\times$ the star formation rate of the average full
resolution clump.  The increase in star formation rate is larger for blurred clumps that are
associated with more than one clump.  These are plausible results,
since the blurred clump luminosity should be very close the sum of the
full resolution clumps, ie for a blurred clump $L_{blurred}\sim \Sigma
L_{full} \sim N_{clump} <L_{full}>\sim 2-5 <L_{full}>$. The largest
enhancements to the blurred clump star formation rate occur in those clumps located in
rings, again where clumps are more crowded and emission from the ring
can also be included in the blurred measurement.
\begin{figure*}
\begin{center}
\includegraphics[width=0.99\textwidth]{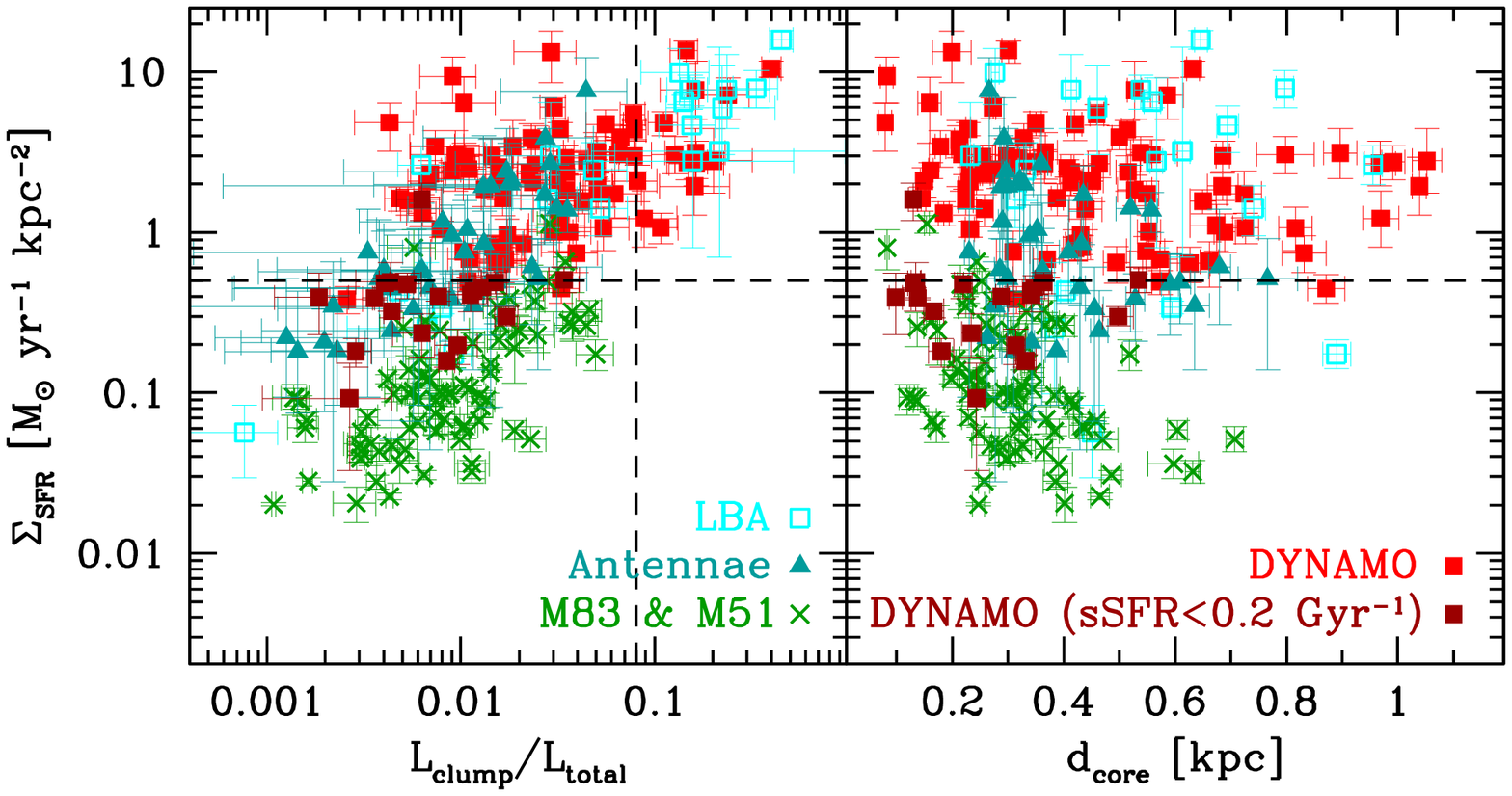}
\end{center}
\caption{ Here we compare the surface density of clumps to the
  clump-to-total H$\alpha$ ratio (left) and the core diameter of
  clumps (right panel). Comparison of clump properties of DYNAMO
  clumps to clumps in nearby star forming disks M83 \& M51, and the
  Antenna advanced-stage merging system. Note that the
  images for , Antenna, M~83 and M~51 have been
  blurred to match the resolution of a galaxy observed at $z=0.1$
  with ACS/WFC. Symbol colors and shapes are as described in the
  key. Symbols for DYNAMO clumps are same as described in
  Fig.~\ref{fig:radlumhi}. In our data set the star formation rate surface density
  provides a single paramter that can clearly distinguish clumps in
  galaxies like DYNAMO and LBA systems, to those in more typical
  $z\sim 0$ galaxies like A04-3, M~51 and M~83.
}
\label{fig:localcomp}
\end{figure*}

For blurred DYNAMO clumps the surface brightness can be as much as
20$\times$ fainter than when measured on the full resolution maps
(right panel of Fig.~\ref{fig:blur_effects}). The typical effect on
$\Sigma_{SFR}$ is to decrease the surface brightness by roughly a
factor of $6\times$ the average surface brightness of corresponding
clumps in the full resolution map.

In short our data suggests the clustered nature of clumps appears to
exacerbate the effects of resolution degradation on clump measurements
in blurred DYNAMO maps. Clump clustering enhances the measured size of
clumps significantly more than simply the blurring due to the larger
physical size of the beam. 

\section{The star formation rate surface density of giant clumps compared to
star forming regions in nearby galaxies}

In massive nearby disk galaxies the star formation rate surface density ranges between
$10^{-4}-10^{-1}$~M$_{\sun}$~yr$^{-1}$~kpc$^{-2}$
\citep[e.g.][]{bigiel2008,leroy2008,rahman2012}. At $z\sim 0$ very
high star formation rate surface densities are only found in extreme environments such
as advanced stage mergers \citep{sanders1996} or disk galaxy centers
with high stellar surface densities \citep{fisher2013}.  However we
report that the star formation rate surface density of DYNAMO clumps in the main body
of a set of rotating, exponential disks that is an order of magnitude
higher than what is typical of star forming regions in disk galaxies.
This is consistent with the star formation rate surface density of clumps in lensed
galaxies \citep{livermore2012}. In Fig.~\ref{fig:localcomp} we show
that there is very little overlap between the star formation rate
surface density of clumps and that of star forming regions in nearby
spirals. 

The vertical line in the left panel of Fig.~\ref{fig:localcomp}
represents an H$\alpha$ clump that is 8\% of the total light of the
galaxy. 
There is a strong correlation with the surface density of the star
forming regions and the clumpiness of the galaxy. Combining the sample
of clumpy DYNAMO galaxies and LBA galaxies we find that in the same
sample only 7 of 103 clumps have $\Sigma_{SFR}<
0.5$~M$_{\odot}$~yr$^{-1}$~kpc$^{-2}$.  Conversely in the sample of
more modestly star forming spiral galaxies (M~83, M~51, C~13-1 and
A~04-3) we find that no star forming region in these galaxies is
brighter than $L_{clump}/L_{total}>8$\%, and only 5 out of 101 clumps
have $\Sigma_{SFR}> 0.5$~M$_{\odot}$~yr$^{-1}$~kpc$^{-2}$.  Using a
dividing line of $\Sigma_{SFR}= 0.5$~M$_{\odot}$~yr$^{-1}$~kpc$^{-2}$
to identify giant clumps is therefore accurate to 95\%.

In the right panel of Fig.~\ref{fig:localcomp} we also compare the
clump size ($d_{core}$) to the star formation rate surface
density. DYNAMO clumps are not systematically smaller than star
forming regions in nearby disks. In fact the opposite is true, DYNAMO
clumps on average are larger, with significant overlap.

Similar to local disks we also measure clump properties in The Antenna
advanced stage merger system. 
None of the clumps in the Antennae are larger than the 8\% luminosity
threshold set by the nearby disks, nonetheless the Antennae system has
high $\Sigma_{SFR}$ (median 0.6
~M$_{\odot}$~yr$^{-1}$~kpc$^{-2}$). The distribution of $\Sigma_{SFR}$
in the Antennae system bridges the divide between local spirals and
turbulent disks. The average core diameter of Antennae clumps is
400~pc, similar size to those in DYNAMO galaxies. The clumps in
DYNAMO mergers are somewhat larger than in clumpy disks, in DYNAMO
mergers $<d_{core}>\sim 0.6$~kpc, compared to $0.4$~kpc for clumps in
clumpy disks. There is, however, almost no difference in the typical
clump star formation rate surface density in DYNAMO mergers compared to those in disks
$<\Sigma_{SFR}> \sim 2.5$~M$_{\odot}$~yr$^{-1}$~kpc$^{-2}$ in DYNAMO
mergers, compared to 2.9 ~M$_{\odot}$~yr$^{-1}$~kpc$^{-2}$ in clumpy
DYNAMO disks. We do not find any significant difference in the
relationship between the size and luminosity of clumps in DYNAMO
mergers vs those in clumpy disks.

In short, it appears that $\Sigma_{SFR}$ may provide a simple, single
parameter that can distinguish the star formation in massive star
forming clumps from that of star forming regions more commonly found
in $z\sim 0$ disk galaxies. Combining $\Sigma_{SFR}$ with the maximum
clump-to-total flux ratio further seperates all systems that are more
typical of the low redshift Universe (Antenna, M83, and M51) from
those that are interpretted as being more similar to galaxies of the
high-$z$ Universe (DYNAMO and LBAs).

\section{Summary \& Discussion of Results}

We have analyzed the properties of H$\alpha$ emitting regions in a
sample of galaxies that are very well matched in properties to
$z\sim1$ main sequence galaxies. 

\subsection{Similarity of DYNAMO to high-z clumpy
  galaxies} 
When blurred to match the resolution of a galaxy at $z\sim 2$ DYNAMO
galaxies appear indistinguishable from galaxies observed in high
redshift surveys
\citep[eg.~][]{genzel2011,wisnioski2012,swinbank2012}. In blurred
maps, 8 of 10 DYNAMO galaxies satisfy the criteria from \cite{guo2015}
for a ``clumpy'' galaxy (that being at least one clump contains 8\% of
the light, and is not co-located with the galaxy center). Clumps in
these blurred DYNAMO maps have a similar distribution of both
size and luminosity relative to the host galaxy to those in $z\sim 1-2$ galaxies
(Fig.~\ref{fig:z2props}). We therefore conclude that DYNAMO galaxies
are indeed ``clumpy'' galaxies. We find that DYNAMO galaxies with
increasing $\sigma/V$ and specific star formation rate contain clumps that are a
larger fraction of the total H$\alpha$ luminosity of the galaxy.

This result is consistent with results from \cite{green2013} and
\cite{fisher2014} that DYNAMO galaxies have rotating kinematics with
high internal velocity dispersions ($\sigma_{H\alpha}\sim
30-60$~km~s$^{-1}$), and are rich in molecular gas ($f_{gas}\sim
30\%)$. Recently \cite{obreschkow2015} show that DYNAMO disk galaxies
are significantly lower angular momentum systems than more typical
$z\sim 0$ disks, which they argue corresponds to the expectation for
similar mass galaxies at $z\sim 2$. Thus it appears that in many
critical ways DYNAMO galaxies are very similar to $z\sim 1-2$
main-sequence disk galaxies.  A straightforward interpretation of
these results is that if galaxy is extremely gas rich disk and that
gas is turbulent, then independent of the galaxy's redshift the
internal physics of gas and subsequent star formation remains very
similar.

\subsection{Effect of clump clustering on clump properties measured at
  $\sim 1$~kpc resolution}
When studied at 100~pc resolution, the sizes of DYNAMO clumps are
similar to those in lensed galaxies
\citep{jones2010,livermore2012,livermore2015} as well as star forming
regions in LBAs (Fig.~\ref{fig:radlumhi} and
\citealp{overzier2009}). In all studies to date of gas rich and/or
turbulent disk galaxies that have resolution below $\sim$200~pc the
size of clumps is found to be a few hundred parsecs.

We find, using blurred DYNAMO maps, that the combination of the both
poorer resolution and the spatial distribution of clumps in the galaxy
exacerbates the effect on the properties of clumps when viewed with
FWHM$\sim 1-2$~kpc.  As we illustrate in Fig.~\ref{fig:g13_1} clumps
that are clustered together in high resolution DYNAMO {\em HST} H$\alpha$
maps appear as a single much larger clump in maps that have been
blurred to match $z\sim 2$ resolution. 

In the DYNAMO sample clump clustering affects the basic properties of
clumps in systematic ways. The systematic affects on clump properties
are shown in
Figs.~\ref{fig:deltaR}, and \ref{fig:blur_effects}. \\
$\bullet$ {\em Size of clumps is increased more than simply the size
  of the beam.} We find that excessively large clumps ($d_{core}\sim
2-3$~kpc) in blurred DYNAMO maps correspond to larger numbers of
DYNAMO clumps in the nominal $z\sim 0.1$ resolution maps. \\ \\
$\bullet$ {\em Number of clumps are significantly reduced:} When
multiple clumps are clustered together in a higher resolution map and
then imaged at lower resolution those clumps are counted as a single
clump.  Clump cluster systematically acts to reduce measured number of
clumps from $\sim 7-12$ clumps per galaxy to $\sim 1 -3$ clumps per
galaxy at the lower
($FWHM\sim 1-2$~kpc) resolution. \\ \\
$\bullet$ {\em Clump clustering increases star formation rate:} On average the
increase in star formation rate due to resolution is found to be a
factor of a few. Some studies assume that resolution generates
 an order of magnitude difference in the baryonic mass
\citep[e.g.][]{tamburello2015}. These studies cite the difference in
clump masses in measured lensed galaxies from those in unlensed
galaxies as evidence for this decrease. However the total mass or star formation rate
of the targeted galaxy must be considered. As we show in
Fig.~\ref{fig:ms}, lensed samples on average target lower mass and
lower star formation rate galaxies compared to both DYNAMO and samples of unlensed
high-$z$ clumpy galaxies. If we assume that the mass of the clump is
mainly gas and that the gas mass scales with the star formation rate, then our results
suggest such a large difference is possible but not the rule. This
result is consistent with a straightforward logic: if a few clumps are
clustered together the blurred clump will be the sum of the flux from
the clumps plus any lower emission flux that surrounds them. Since the
clumps are by definition brighter than the surrounding region, the
increase in flux will be $L_{blur}/<L_{resolved}> \sim N_{clumps}$.
It is very unlikely that this effect could lead to an order of
magnitude increase in flux or thus mass of blurred clumps.

We find that clumps located in rings are most affected by the loss in
resolution, creating a typical increase in blurred clump star formation rate of $\sim
5\times$ that of the unblurred clumps. This is likely because both
clumps are clustered more in rings, and the H$\alpha$ emission in the
ring likely is combined into the blurred clump flux. \\ \\
$\bullet$ {\em Clump star formation rate surface density is strongly affected by
  resolution} The larger sizes of blurred clumps creates
systematically lower star formation rate surface densities by as much as an order of
magnitude, and also significantly fewer numbers of clumps per galaxy
(Fig.~\ref{fig:blur_effects}).  The star formation rate surface density of full
resolution clump measurements ranges 2-20$\times$, and is typically a
factor of 6$\times$ that of the blurred clump.  If we consider the
implication of this on the results of high-redshift galaxies, where
impressivelly high star formation rate surface densities for clumps of
1-10~M$_{\odot}$~yr$^{-1}$~kpc$^{-2}$ are reported with a beam
FWHM$\sim 1.5-2$~kpc \citep{genzel2011}, an average factor
of 6$\times$ increase to these star formation rate surface densities the properties
are even more extreme.

\cite{meurer1997} find that the star burst intensity of UV bright
galaxies is limited, as indicated by the absence of galaxies with
greater than $\Sigma_{SFR}(limit)\sim
25$~M$_{\odot}$~yr$^{-1}$~kpc$^{-2}$. The implication of such a limit
is that some physical mechanism exists which regulates the star
formation, and this mechanism is Universal. The derived $\Sigma_{SFR}$
of clumps in the samples with study is consistent with this limit
\citep{genzel2011}. However, if these measured $\Sigma_{SFR}$ are
artificially low due to resolution effects, then clumps in $z\sim 1-2$
main-sequence galaxies may violate this limit, similar to the handful
of super-intense starburst in observed in extreme galaxies
\citep[e.g.~][]{swinbank2010,hodge2015}.

\subsection{Star formation rate surface density of clumps in turbulent disks } 
We show in Fig.~\ref{fig:localcomp} that the star formation rate surface density of
DYNAMO clumps is $\Sigma_{SFR}\sim
1-10$~M$_{\odot}$~yr$^{-1}$~kpc$^{-2}$. This is orders of magnitude
higher than what is found in the star forming regions of nearby spiral
galaxies \citep[e.g.][]{bigiel2008,leroy2008,rahman2012}. In the local
Universe the only other place to find such high star formation rate surface densities
are shocks in galaxy-galaxy mergers
\citep{wei2012,zaragoza2014} or galaxy centers
\citep{fisher2013}.

In Fig.~\ref{fig:localcomp} we show that the star formation rate
surface density offers a simple, physically motivated method to
distinguish massive star forming clumps, characteristic of star
formation in massive $z\sim 1-2$ disk galaxies, from star formation
that is more typical of low gas fraction disks of the nearby Universe.
The dividing line of $\Sigma_{SFR}\sim
0.5$~M$_{\odot}$~yr$^{-1}$~kpc$^{-2}$ is shown to be a good way to
distinguish massive star forming clumps from more typical star forming
regions of $z\sim 0$ spiral galaxies.

Following arguments from the \cite{larson1981} relations \citep[also
see arguements in]{elmegreen1989} the pressure increases in molecular
clouds as the square of the surface density of the molecular gas,
$P_{ext} \propto \Sigma_{mol}^2$. Clumps in DYNAMO galaxies are shown
to have $\Sigma_{SFR}$ that are $\sim 500\times$ that of star forming
regions in nearby spirals. It is likely that the gas mass surface
density is similarly high in clumps. We convert the star formation
rate surface density to gas mass surface density using the global
scaling for DYNAMO disks in \cite{fisher2014} of find $M_{mol}/SFR
\sim 5\times10^8$~yr. The pressure in the massive star forming clumps
in a DYNAMO galaxy is therefore $10^4 \times$ that found in star
forming regions of nearby spirals. These high pressures are only observed in
the most extreme clouds in the center of the Milky Way
\citep{rathborne2014}, and high-$z$ galaxies
\citep{swinbank2010,swinbank2015arxiv}. It remains unclear if the
conditions in these environments are consistent with models of star
formation \citep[eg.~][]{krumholz2005}. These systems therefore
warrant more work with facilities such as ALMA to accurately describe
the star formation efficiency, and ratios of dense gas gas inside the
star forming regions, and also to understand what drives the turbulent
gas that creates these clumps. $\\ $

In this work we have again shown that
 the DYNAMO sample is an excellent laboratory to study the conditions
of galaxies at $z\sim1-3$, with which they are consistent. In future
work we will compare 
expectations from the violent disk instability model to the clump
properties of DYNAMO galaxies (Fisher et al. {\em in prep}), investigate
systematic correlations fo DYNAMO dust and gas properties in galaxies
(Bassett et al. {\em submitted}, White et al. {\em in prep}), and study
the properties of clumps in stellar continuum (Fisher et al. {\em in
  prep}).  

\section*{Acknowledgements}
DBF acknolwedges support from Australian Research Council (ARC)
Discovery Program (DP) grant DP130101460.  Support for this project is
provided in part by the Victorian Department of State Development,
Business and Innovation through the Victorian International Research
Scholarship (VIRS).  
\begin{table*}
\centering
\label{tab:fit_data}
\begin{tabular}{cccc}
\hline
Galaxy  &  $d_{core}$       &   $L_{H\alpha}$           &  SFR      \\
        &   (pc)            &    ($10^{40}$ erg s$^{-1}$) & ($M_{\odot}$ yr$^{-1}$) \\
\hline
G20-2	&	547	$\pm$	34	&	4.95	$\pm$	1.67	&	0.27	\\
G20-2	&	968	$\pm$	28	&	54.03	$\pm$	42.98	&	2.99	\\
G20-2	&	896	$\pm$	30	&	53.68	$\pm$	37.63	&	2.97	\\
G20-2	&	159	$\pm$	21	&	3.43	$\pm$	1.50	&	0.19	\\
G20-2	&	1052	$\pm$	26	&	66.19	$\pm$	61.92	&	3.66	\\
G20-2	&	798	$\pm$	51	&	41.93	$\pm$	23.83	&	2.32	\\
G20-2	&	305	$\pm$	9	&	18.16	$\pm$	5.39	&	1.00	\\
G20-2	&	217	$\pm$	59	&	3.09	$\pm$	1.31	&	0.17	\\
G20-2	&	674	$\pm$	47	&	10.66	$\pm$	5.13	&	0.59	\\
G20-2	&	650	$\pm$	23	&	13.99	$\pm$	5.54	&	0.77	\\
G20-2	&	626	$\pm$	37	&	5.39	$\pm$	2.16	&	0.30	\\
G13-1	&	586	$\pm$	29	&	52.82	$\pm$	19.33	&	2.92	\\
G13-1	&	632	$\pm$	12	&	88.60	$\pm$	26.35	&	4.90	\\
G13-1	&	555	$\pm$	18	&	12.63	$\pm$	3.74	&	0.70	\\
G13-1	&	549	$\pm$	21	&	11.30	$\pm$	3.21	&	0.62	\\
G13-1	&	871	$\pm$	34	&	7.29	$\pm$	3.85	&	0.40	\\
G13-1	&	430	$\pm$	22	&	3.84	$\pm$	1.08	&	0.21	\\
G13-1	&	553	$\pm$	21	&	6.66	$\pm$	2.19	&	0.37	\\
G13-1	&	495	$\pm$	24	&	3.41	$\pm$	1.06	&	0.19	\\
G13-1	&	431	$\pm$	17	&	3.25	$\pm$	0.71	&	0.18	\\
G14-1	&	727	$\pm$	23	&	12.10	$\pm$	4.52	&	0.67	\\
G14-1	&	407	$\pm$	8	&	38.34	$\pm$	6.45	&	2.12	\\
G14-1	&	452	$\pm$	19	&	9.18	$\pm$	2.60	&	0.51	\\
G14-1	&	350	$\pm$	10	&	12.56	$\pm$	2.43	&	0.69	\\
G08-5	&	406	$\pm$	12	&	8.99	$\pm$	1.54	&	0.50	\\
G08-5	&	301	$\pm$	12	&	26.27	$\pm$	4.69	&	1.45	\\
G08-5	&	404	$\pm$	19	&	18.64	$\pm$	4.31	&	1.03	\\
G08-5	&	367	$\pm$	21	&	9.04	$\pm$	3.07	&	0.50	\\
G08-5	&	212	$\pm$	13	&	3.35	$\pm$	1.48	&	0.19	\\
G08-5	&	421	$\pm$	21	&	2.99	$\pm$	0.62	&	0.17	\\
G08-5	&	414	$\pm$	22	&	8.78	$\pm$	2.71	&	0.49	\\
G08-5	&	318	$\pm$	21	&	6.28	$\pm$	1.39	&	0.35	\\
G08-5	&	371	$\pm$	26	&	2.00	$\pm$	0.49	&	0.11	\\
G08-5	&	439	$\pm$	29	&	5.77	$\pm$	1.67	&	0.32	\\
H10-2	&	831	$\pm$	40	&	11.05	$\pm$	6.58	&	0.61	\\
H10-2	&	573	$\pm$	31	&	4.76	$\pm$	1.82	&	0.26	\\
H10-2	&	1039	$\pm$	13	&	44.94	$\pm$	39.62	&	2.49	\\
H10-2	&	527	$\pm$	7	&	45.37	$\pm$	13.53	&	2.51	\\
H10-2	&	445	$\pm$	16	&	8.81	$\pm$	2.55	&	0.49	\\
H10-2	&	969	$\pm$	32	&	24.71	$\pm$	19.39	&	1.37	\\
H10-2	&	816	$\pm$	30	&	15.24	$\pm$	8.94	&	0.84	\\
H10-2	&	689	$\pm$	19	&	10.14	$\pm$	4.70	&	0.56	\\
G04-1	&	233	$\pm$	17	&	2.68	$\pm$	0.57	&	0.15	\\
G04-1	&	254	$\pm$	19	&	4.10	$\pm$	0.76	&	0.23	\\
G04-1	&	661	$\pm$	17	&	6.23	$\pm$	2.43	&	0.34	\\
G04-1	&	284	$\pm$	14	&	4.23	$\pm$	1.08	&	0.23	\\
G04-1	&	538	$\pm$	18	&	19.50	$\pm$	5.81	&	1.08	\\
G04-1	&	685	$\pm$	27	&	19.50	$\pm$	6.35	&	1.08	\\
G04-1	&	992	$\pm$	21	&	57.25	$\pm$	31.38	&	3.17	\\
G04-1	&	686	$\pm$	12	&	30.23	$\pm$	10.24	&	1.67	\\
G04-1	&	297	$\pm$	19	&	5.58	$\pm$	1.59	&	0.31	\\
G04-1	&	199	$\pm$	35	&	11.27	$\pm$	4.11	&	0.62	\\
G04-1	&	464	$\pm$	20	&	12.22	$\pm$	2.95	&	0.68	\\
G04-1	&	328	$\pm$	28	&	8.90	$\pm$	2.21	&	0.49	\\
G04-1	&	515	$\pm$	22	&	13.44	$\pm$	3.98	&	0.74	\\

\hline
\end{tabular}
\end{table*}
\begin{table*}
\centering
\label{tab:fit_data}
\begin{tabular}{cccc}
\hline
Galaxy  &  $d_{core}$       &   $L_{H\alpha}$           &  SFR      \\
        &   (pc)            &    ($10^{40}$ erg s$^{-1}$) & ($M_{\odot}$ yr$^{-1}$) \\
\hline
D13-5	&	573	$\pm$	15	&	3.44	$\pm$	1.50	&	0.19	\\
D13-5	&	315	$\pm$	14	&	0.82	$\pm$	0.19	&	0.05	\\
D13-5	&	420	$\pm$	13	&	17.67	$\pm$	5.55	&	0.98	\\
D13-5	&	528	$\pm$	31	&	11.03	$\pm$	4.42	&	0.61	\\
D13-5	&	514	$\pm$	11	&	24.79	$\pm$	8.97	&	1.37	\\
D13-5	&	500	$\pm$	13	&	21.04	$\pm$	7.55	&	1.16	\\
D13-5	&	558	$\pm$	46	&	20.58	$\pm$	8.99	&	1.14	\\
D13-5	&	459	$\pm$	12	&	24.54	$\pm$	7.66	&	1.36	\\
D13-5	&	724	$\pm$	31	&	19.63	$\pm$	13.14	&	1.09	\\
D13-5	&	119	$\pm$	6	&	5.39	$\pm$	1.00	&	0.30	\\
D13-5	&	365	$\pm$	13	&	7.66	$\pm$	2.28	&	0.42	\\
D13-5	&	412	$\pm$	18	&	7.41	$\pm$	2.10	&	0.41	\\
D13-5	&	303	$\pm$	18	&	4.96	$\pm$	1.40	&	0.27	\\
D13-5	&	388	$\pm$	11	&	5.21	$\pm$	1.56	&	0.29	\\
D13-5	&	179	$\pm$	10	&	2.33	$\pm$	0.63	&	0.13	\\
D13-5	&	273	$\pm$	5	&	9.51	$\pm$	1.82	&	0.53	\\
D13-5	&	258	$\pm$	15	&	1.99	$\pm$	0.48	&	0.11	\\
D13-5	&	224	$\pm$	13	&	1.70	$\pm$	0.35    &	0.09\\	
D15-3	&	149	$\pm$	7	&	1.01	$\pm$	0.22	&	0.06	\\
D15-3	&	184	$\pm$	9	&	0.95	$\pm$	0.19	&	0.05	\\
D15-3	&	82	$\pm$	11	&	1.36	$\pm$	0.44	&	0.08	\\
D15-3	&	339	$\pm$	19	&	6.73	$\pm$	1.94	&	0.37	\\
D15-3	&	213	$\pm$	9	&	3.63	$\pm$	0.90	&	0.20	\\
D15-3	&	79	$\pm$	12	&	0.64	$\pm$	0.22	&	0.04	\\
D15-3	&	224	$\pm$	13	&	1.99	$\pm$	0.51	&	0.11	\\
D15-3	&	83	$\pm$	7	&	1.31	$\pm$	0.35	&	0.07	\\
D15-3	&	231	$\pm$	13	&	2.37	$\pm$	0.62	&	0.13	\\
D15-3	&	261	$\pm$	7	&	3.38	$\pm$	0.68	&	0.19	\\
D15-3	&	309	$\pm$	30	&	1.54	$\pm$	0.59	&	0.09	\\
D15-3	&	420	$\pm$	17	&	3.18	$\pm$	1.30	&	0.18	\\
D15-3	&	161	$\pm$	4	&	1.34	$\pm$	0.18	&	0.07	\\
D15-3	&	232	$\pm$	8	&	1.19	$\pm$	0.20	&	0.07	\\
D15-3	&	228	$\pm$	4	&	4.88	$\pm$	0.74	&	0.27	\\
D15-3	&	144	$\pm$	8	&	0.73	$\pm$	0.14	&	0.04	\\
C13-1	&	535	$\pm$	14	&	3.12	$\pm$	1.54	&	0.17	\\
C13-1	&	364	$\pm$	16	&	1.37	$\pm$	0.37	&	0.08	\\
C13-1	&	288	$\pm$	10	&	0.70	$\pm$	0.12	&	0.04	\\
C13-1	&	130	$\pm$	8	&	0.58	$\pm$	0.15	&	0.03	\\
C13-1	&	219	$\pm$	16	&	0.49	$\pm$	0.10	&	0.03	\\
C13-1	&	235	$\pm$	4	&	5.79	$\pm$	0.88	&	0.32	\\
C13-1	&	348	$\pm$	17	&	1.18	$\pm$	0.31	&	0.07	\\
C13-1	&	342	$\pm$	15	&	1.06	$\pm$	0.28	&	0.06	\\
C13-1	&	342	$\pm$	31	&	1.03	$\pm$	0.30	&	0.06	\\
C13-1	&	498	$\pm$	27	&	1.57	$\pm$	0.68	&	0.09	\\
A04-3	&	234	$\pm$	18	&	0.27	$\pm$	0.09	&	0.02	\\
A04-3	&	98	$\pm$	12	&	0.08	$\pm$	0.03	&	0.004\\
A04-3	&	135	$\pm$	9	&	0.19	$\pm$	0.06	&	0.01	\\
A04-3	&	220	$\pm$	10	&	1.03	$\pm$	0.28	&	0.06	\\
A04-3	&	244	$\pm$	16	&	0.12	$\pm$	0.08	&	0.01	\\
A04-3	&	314	$\pm$	17	&	0.42	$\pm$	0.15	&	0.02	\\
A04-3	&	331	$\pm$	19	&	0.37	$\pm$	0.11	&	0.02	\\
A04-3	&	138	$\pm$	13	&	0.16	$\pm$	0.01	&	0.01	\\
A04-3	&	180	$\pm$	23	&	0.13	$\pm$	0.03	&	0.01	\\
A04-3	&	166	$\pm$	18	&	0.19	$\pm$	0.03	&	0.01	\\
A04-3	&	131	$\pm$	11	&	0.18	$\pm$	0.02	&	0.01	\\
\hline
\end{tabular}
\end{table*}


\appendix
\section{Comparison Samples}
\subsection{Comparison sample 1:  High-z galaxies and high-z analogues}

The DYNAMO-{\em HST} data set occupies a unique section of parameter
space. The FWHM of DYNAMO-HST $H\alpha$+[NII] maps is $\sim
0.07-0.12$~arcsec, corresponding to a physical resolution of
$\sim150-300$~pc. 
The physical resolution from observing DYNAMO galaxies is roughly from
$\sim 3-10\times$ finer compared to AO-enabled or {\em HST} observations of
high-$z$ galaxies, similar to what is achieved from lens galaxies,
however unlike lensed galaxies in DYNAMO the improvement in resolution
is in both directions. As we show in Fig.~\ref{fig:ms} the SFR and
stellar mass of DYNAMO galaxies is typically $\sim 10\times$ that of
typical lensed galaxies with clump measurements. The average clump SFR
is larger in galaxies with more star formation \citep{livermore2012}.
A direct comparison of the properties of star forming clumps should
control for total SFR.

Here we summarize the salient features of each of the high-$z$
comparison data sets of clump properties.

{\bf SINS:} The clump measurements are taken from \cite{genzel2011},
which is a subset of the SINS survey \citep{forsterschreiber2009}. As
we show in Fig.~\ref{fig:ms}, the galaxies in the \cite{genzel2011}
sample have the largest star formation rates in our comparison sample,
ranging 30-200~M$_{\odot}$~yr$^{-1}$. Stellar masses of this SINS
subset range 10$^{10}-10^{11}$~M$_{\odot}$. The galaxies in the SINS
clump sample have $z\sim 2.2-2.3$. The observations of the SINS clump
sample were made with SINFONI-AO, achieving typical resolution of
$\sim 0.22$ arcsec, or 1.85~kpc.  \cite{genzel2011} use the width of
Gaussians as a metric of clump sizes, similar to what we use.

{\bf WiggleZ:} The clump properties are taken from
\cite{wisnioski2012}, which is a subset of $z\sim 1.3-1.5$ galaxies in
the WiggleZ Dark Energy survey. The galaxy masses and SFR ($5-10\times
10^{10}$~M$_{\odot}$ and 10-30~M$_{\odot}$~yr$^{-1}$) are consistent
with the $z\sim 1$ main-sequence. Observations of WiggleZ clumps are
made with Keck/OSIRIS-AO and have a resolution of $\sim 0.1$~arcsec,
or 0.85~kpc. \cite{wisnioski2012} uses Gaussian RMS. We therefore
adjust the \cite{wisnioski2012} clumps sizes $R= R_{wis} \time 2\sqrt(2
ln_{}2)$.

{\bf SHiZELS:} The clump measurements for the SHiZELS survey are taken
from \cite{swinbank2012}. Galaxy masses span 0.6-10$\times
10^{10}$~M$_{\odot}$ and SFR span 1-30~$M_{\odot}$~yr$^{-1}$. The
redshifts of 3 SHiZELS galaxies are $z\sim 1.5$, and 1 at $z\sim
2.2$. Observations of these galaxies are made with SINFONI+AO with
approximate spatial resolution of 0.1 arcsec, corresponding to 0.85~kpc.

{\bf Jones + 2010:} These observations are made on a sample of
strongly lensed, star forming galaxies at $z\sim 1.7-3.1$. The masses
range $0.5-2 \times 10^{10}$~M$_{\odot}$ and the SFR range
2-40~M$_{\odot}$~yr$^{-1}$. Observations are made with the Keck/OSIRIS
AO system, obtaining resolution of 0.11 arcsec. The lensing
magnification in these galaxies was typically $\sim 8\times$, which
results in typical resolution one direction of $\sim 0.1$~kpc. Note
however this magnification is only in one direction.

In this paper we determine clump sizes through a Gaussian fitting
technicque (described in section 3).  \cite{jones2010} and
\cite{swinbank2012} both use ``isophotal sizes''. \cite{wisnioski2012}
shows that this method systematically returns larger sizes compared to
the Gaussian fitting methods, but an adjustment factor is not
determined. We therefore use these published sizes as is, with the
caveat that in comparison to our clump sizes, these may be slightly
larger.

{\bf Livermore + 2012: } These observations of clumps are also made on
strongly lensed galaxies in the redshift range $z\sim
0.9-1.5$. Though total masses are not given for this sample, typical
SFR range 0.4-12~M$_{\odot}$~yr$^{-1}$. The observations are made with
narrowband filters on {\em HST}/WFC3, the median resolution in one direction
for the sample
is 0.36~kpc. 

{\bf Livermore + 2015:} These observations of clumps are also made on
strongly lensed galaxies in the redshift range $z\sim
1-4$. For galaxies in this sample the 
SFR ranges 0.8-30~M$_{\odot}$~yr$^{-1}$ and total stellar masses range
0.06 - 0.3$\times 10^{10}$~M$_{\odot}$.The observations are made with
SINFONI IFU on the ESO/VLT, the resolution one direction for the sample
spans 0.1-0.7~kpc.  

The \cite{livermore2015} sample also includes a re-analysis of the
galaxies in \cite{jones2010}. In the interest of reducing systematic
biases, and increases the base of our comparison we opt to use values
from \cite{jones2010}.

\cite{livermore2012} and \cite{livermore2015} uses the sizes of clumps
returned from the software CLUMPFIND. \cite{livermore2012} shows that
sizes produced from CLUMPFIND are systematically 25\% larger than
Gaussian FWHM sizes (which we use in this paper); we therefore adjust
the sizes from CLUMPIND by $R = R_{liv}/1.25$.
  
{\bf Lyman Break Analogs (LBA)} - We also include a sample of Lyman
Break Analog galaxies for comparison. These are not high-redshift
galaxies, however ther have properties that are very similar to high
redshift Lyman Break Galaxies
\citep{heckman2005,ostlin2009,goncalves2010}. We use galaxies
Haro~11, IRAS08339+6517 and NGC~6090 from \cite{heckman2011}
describing LBA galaxies. The {\em HST}
maps we use are presented and discussed in \cite{ostlin2009}. As we
discuss above, LBA systems are more likely to resemble high-$z$
merging galaxies than turbulent disks. 
The LBA sample therefore provides a complimentary
comparison to the DYNAMO sample. We degrade H$\alpha$ maps of the LBA
galaxies to match the resolution of a galaxy at $z\sim 0.1$ that is
observed with ACS/WFC, so that there is a straightforward comparison
to DYNAMO galaxies. The clump properties are derived in the same
procedure as is used for the DYNAMO galaxies.

\subsection{Comparison Sample 2: Nearby Spiral and Merging Galaxies}

As is indicated in Fig.~\ref{fig:ms}, the DYNAMO-{\em HST} sample contains
one galaxy that is selected to be consistent with the $z\sim0.1$ main
sequence. This galaxy, A04-3, has both a modest SFR =
3.4~M$_{\odot}$~yr$^{-1}$ and velocity dispersion $\sigma_{H\alpha} =
10$~km~s$^{-1}$. We choose to observe this galaxy as a control for our
method. We also make similar measurements to clumps properties on
galaxies in the nearby Universe. We choose two well known disks and
one well known merger. To make the comparison as uniform as
possible we select only comparison galaxies with available
H$\alpha$+[NII] maps available in the {\em HST} archive. We then degrade the
resolution of the image to match that of a galaxy observed at $z=0.1$
with ACS/WFC.

The comparison samples include the following galaxies: \\
{\bf Arp 244} - The so-called Antennae galaxy is an advanced stage
merger with a large set of very active star forming regions. We
utilize the the map from \cite{whitmore2010}
targeting the middle, star bursting section of the galaxy. The total SFR
of the Antennae system is $\sim
5-10$~M$_{\odot}$~yr$^{-1}$ \citep{brandl2009},
placing this galaxy at the low end of the
SFRs observed in DYNAMO galaxies. \\
{\bf M~83 and M~51} - These galaxies are chosen to be representative
of $z\sim 0$ star forming disks. The galaxies M~83 and M~51 have
comparable stellar mass to DYNAMO galaxies ($1-3\times
10^{10}$~M$_{\odot}$), however compared to most of the galaxies in the
DYNAMO-{\em HST} sample they have much more modest star formation rates
($\sim 1-2$~M$_{\odot}$~yr$^{-1}$), and also lower molecular gas
fractions ($\sim 1-5$\%).

We convolved the H$\alpha$ maps of M~83, M~51 and Arp~244 with a
symmetric 2-D Gaussian to simulate the resolution they would have if
these well known, nearby galaxies were observed with ACS/WFC
resolution at $z=0.1$.

\section{Surface Photometry of Continuum Maps and Disk/Merger classification}
\begin{figure*}
\begin{center}
\includegraphics[width=0.3\textwidth]{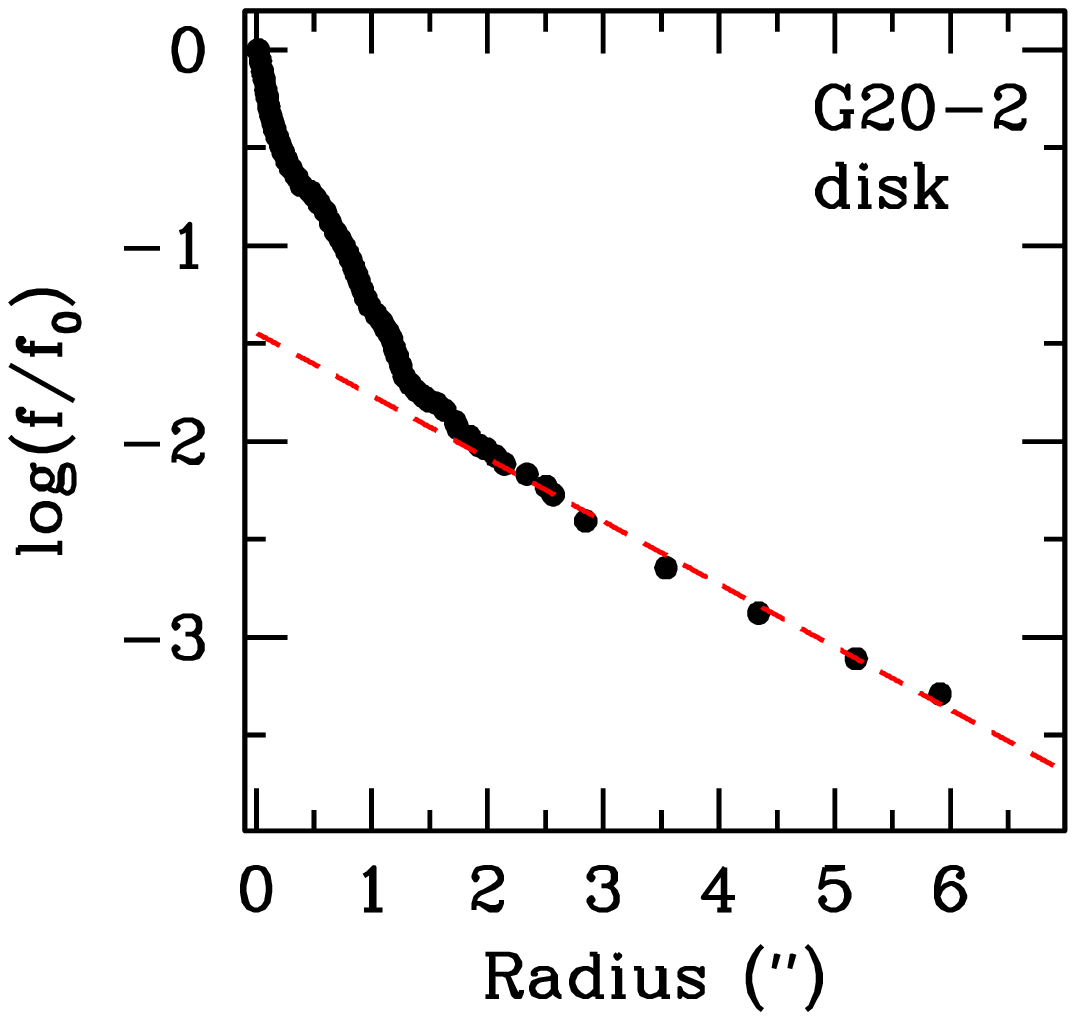}
\includegraphics[width=0.3\textwidth]{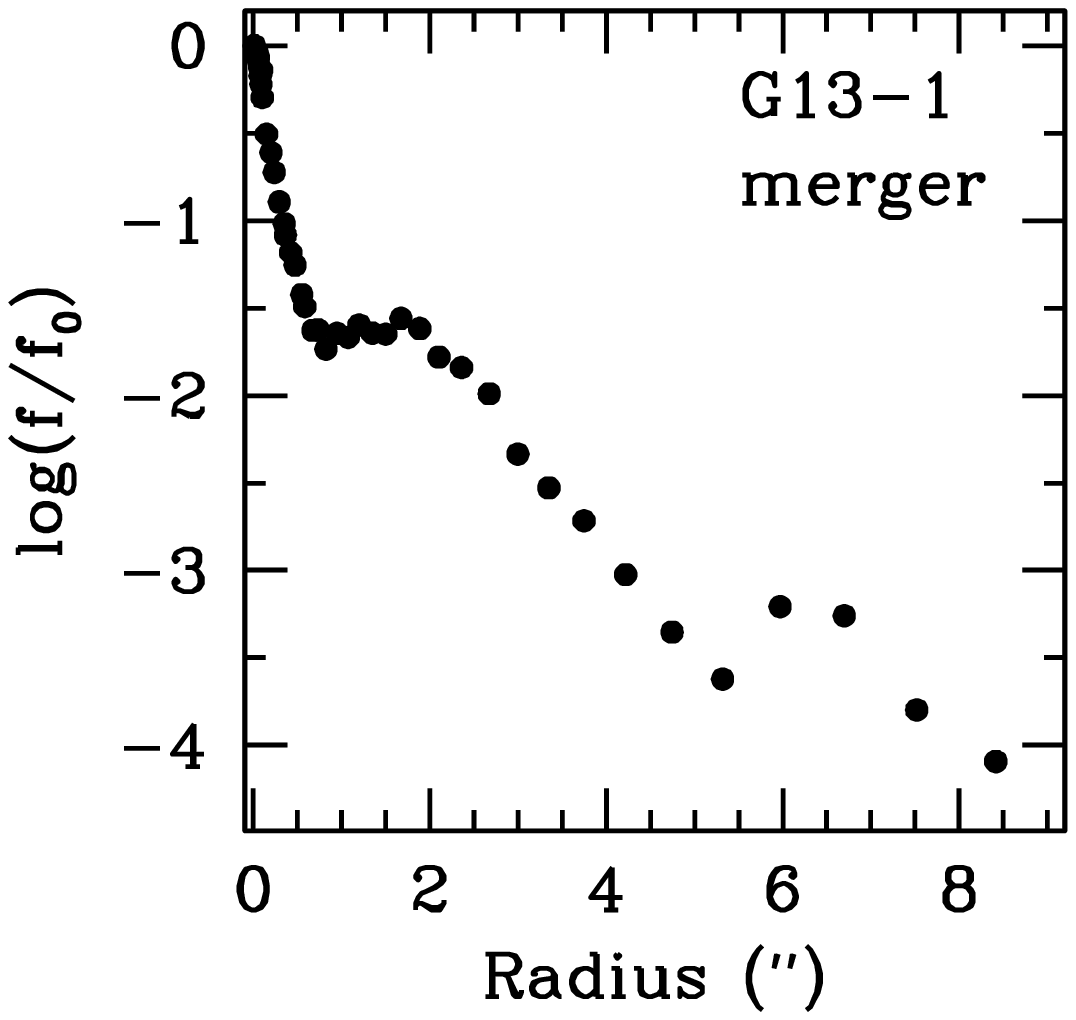}
\includegraphics[width=0.3\textwidth]{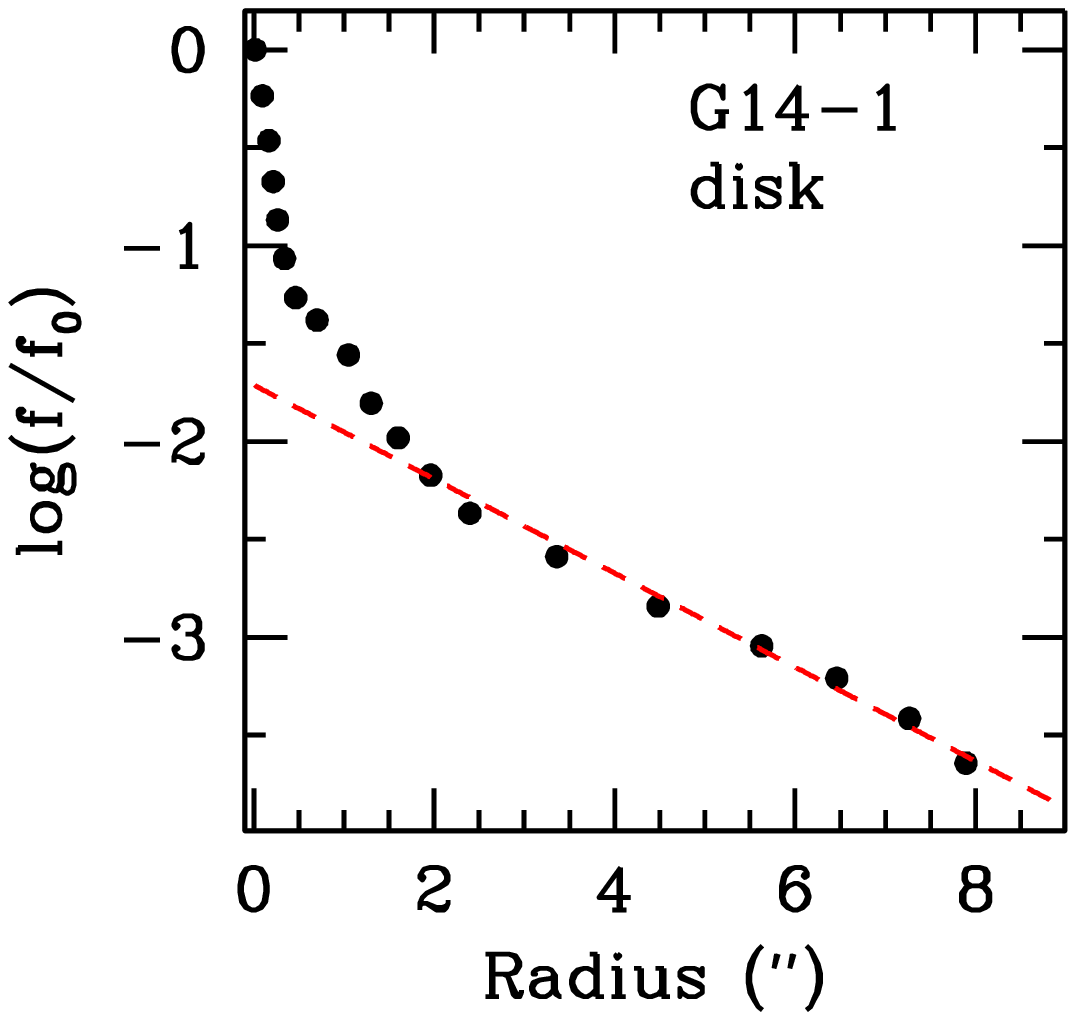} \\
\includegraphics[width=0.3\textwidth]{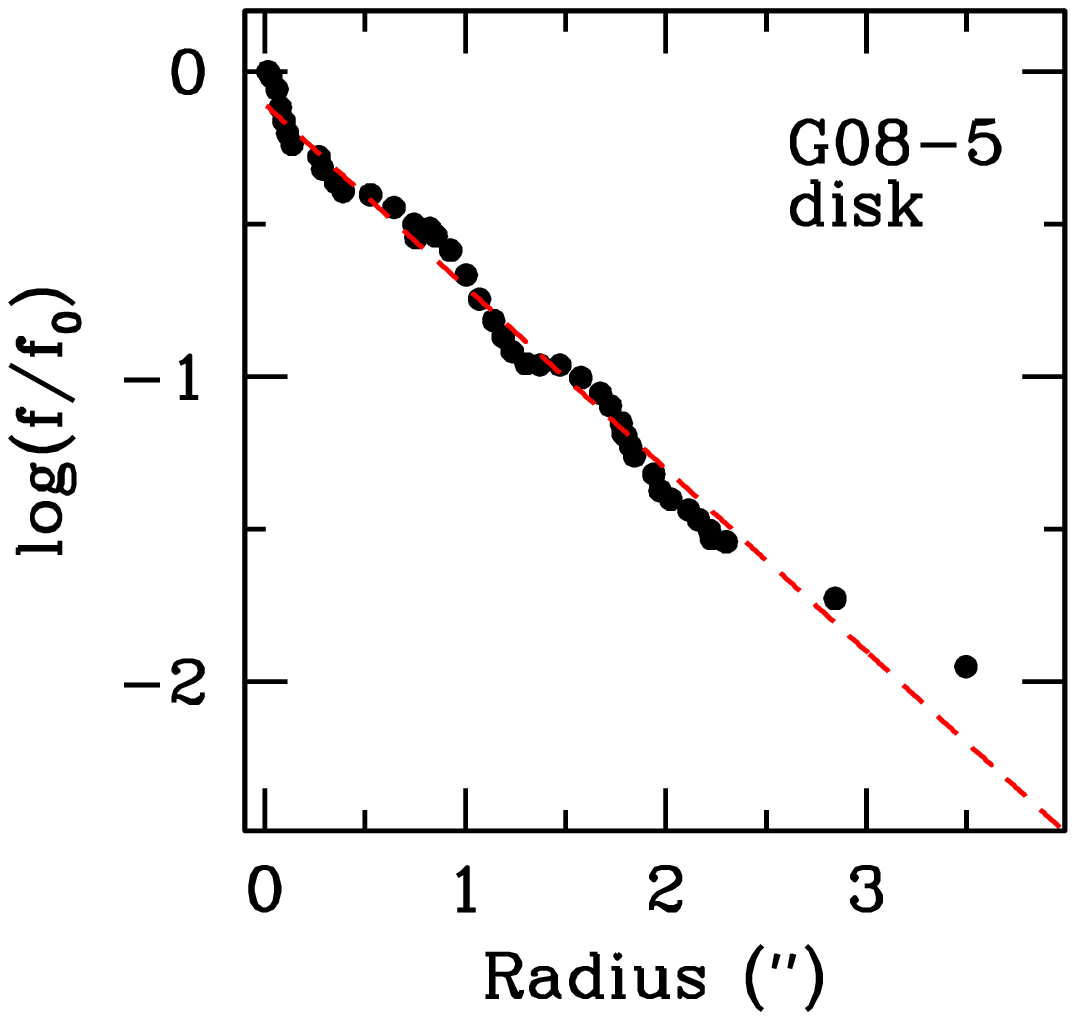}
\includegraphics[width=0.3\textwidth]{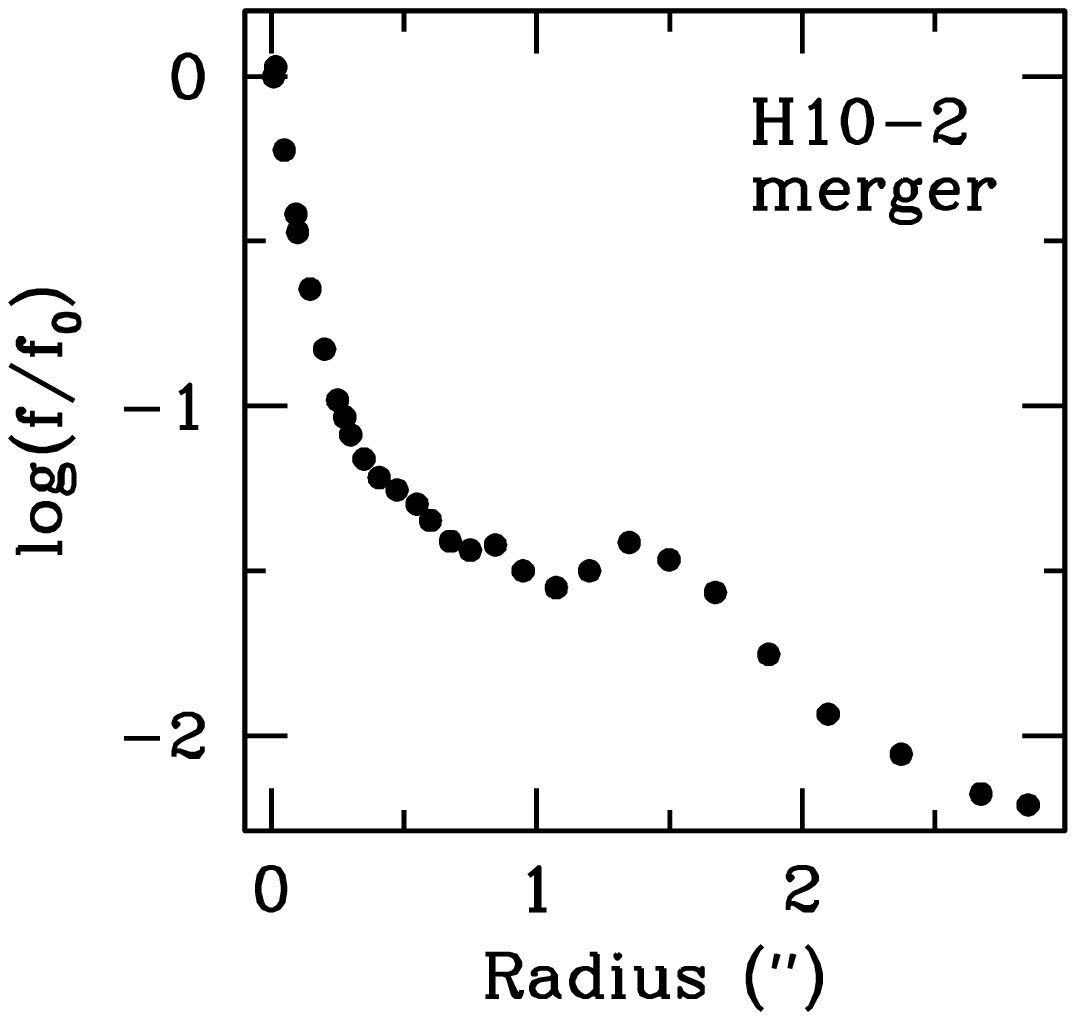}
\includegraphics[width=0.3\textwidth]{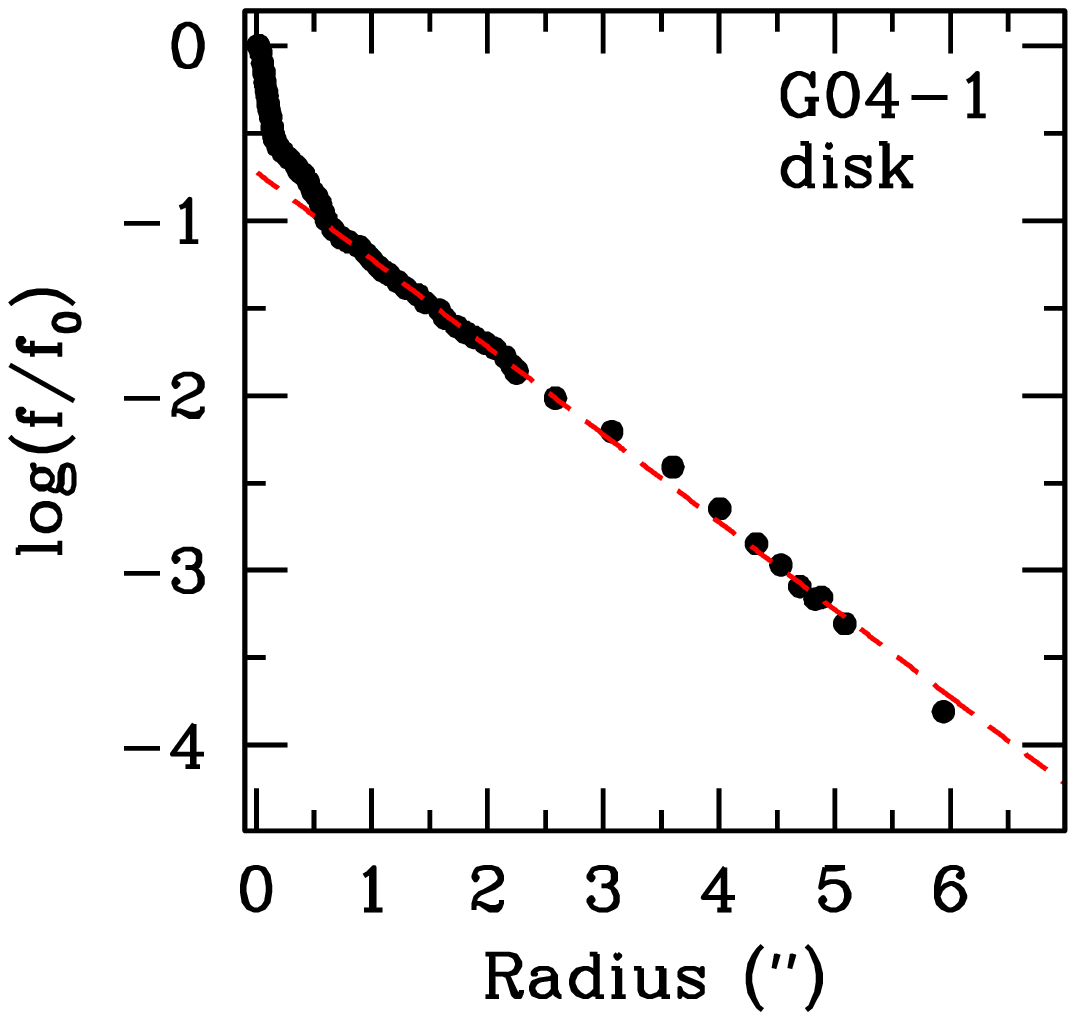} \\
\includegraphics[width=0.3\textwidth]{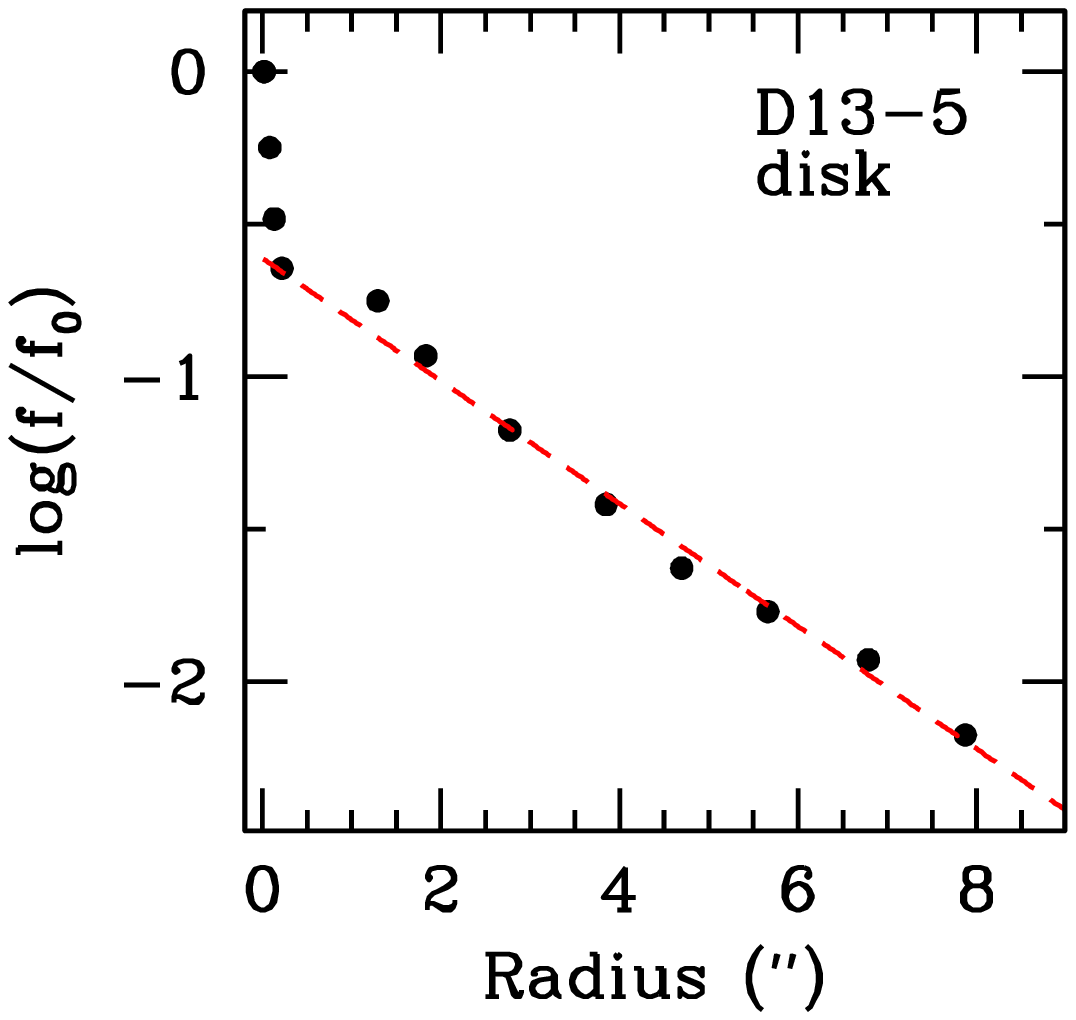}
\includegraphics[width=0.3\textwidth]{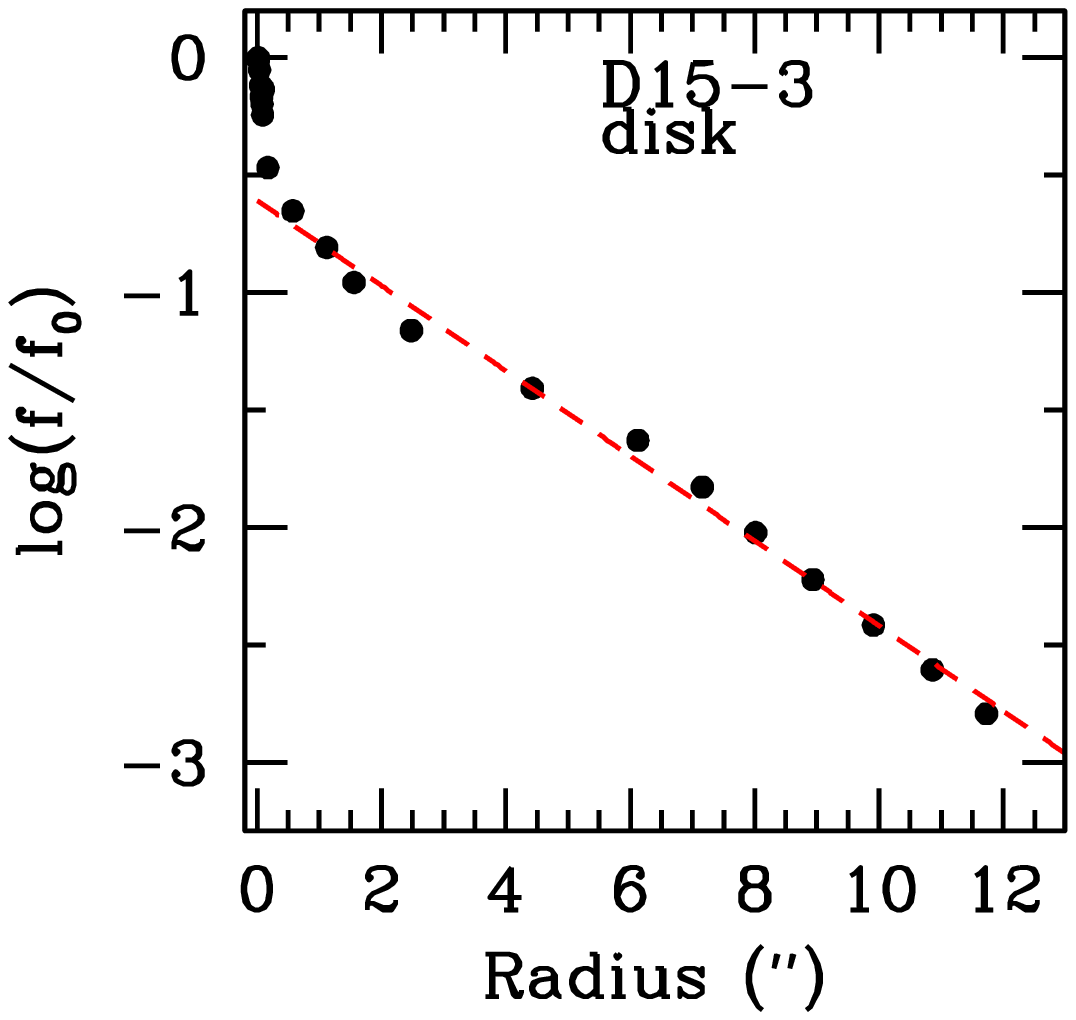}
\includegraphics[width=0.3\textwidth]{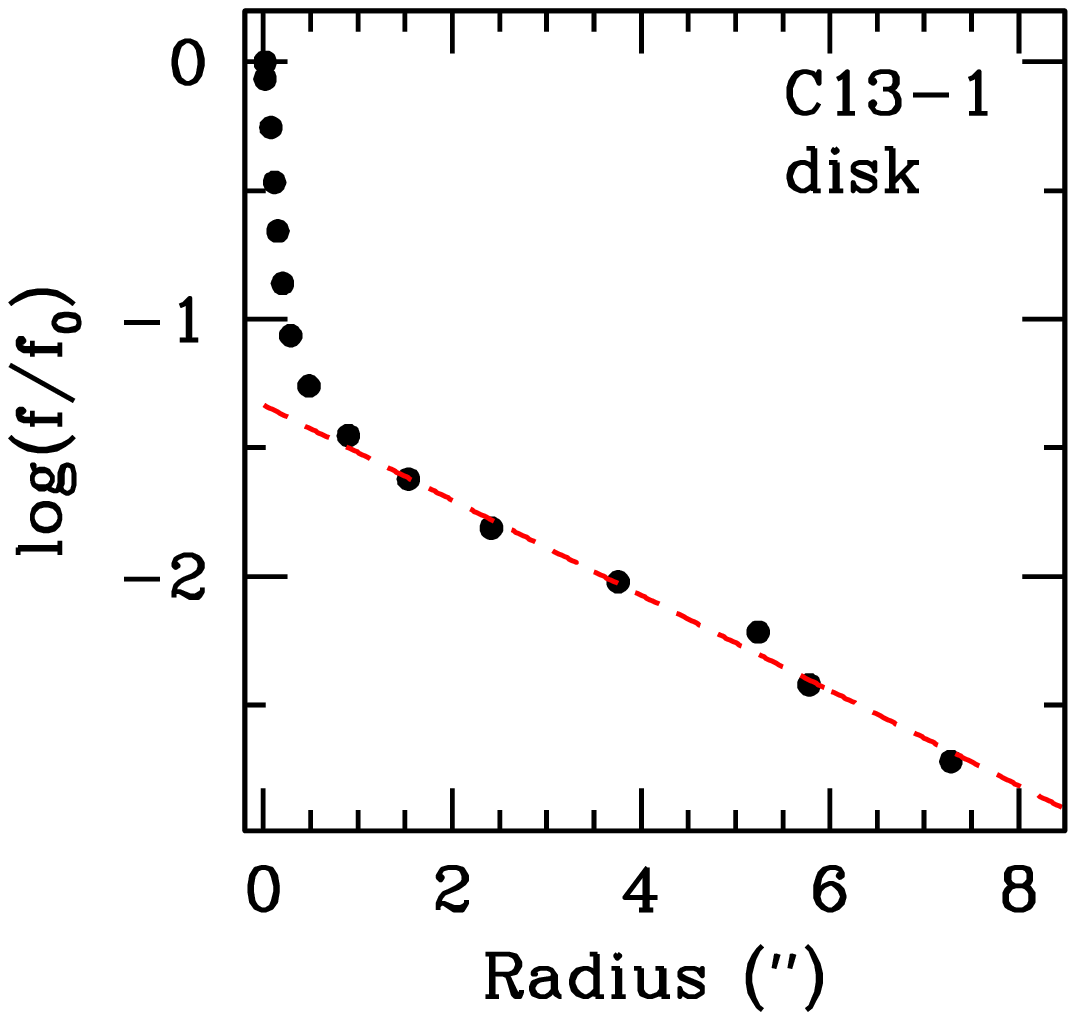} \\
\includegraphics[width=0.3\textwidth]{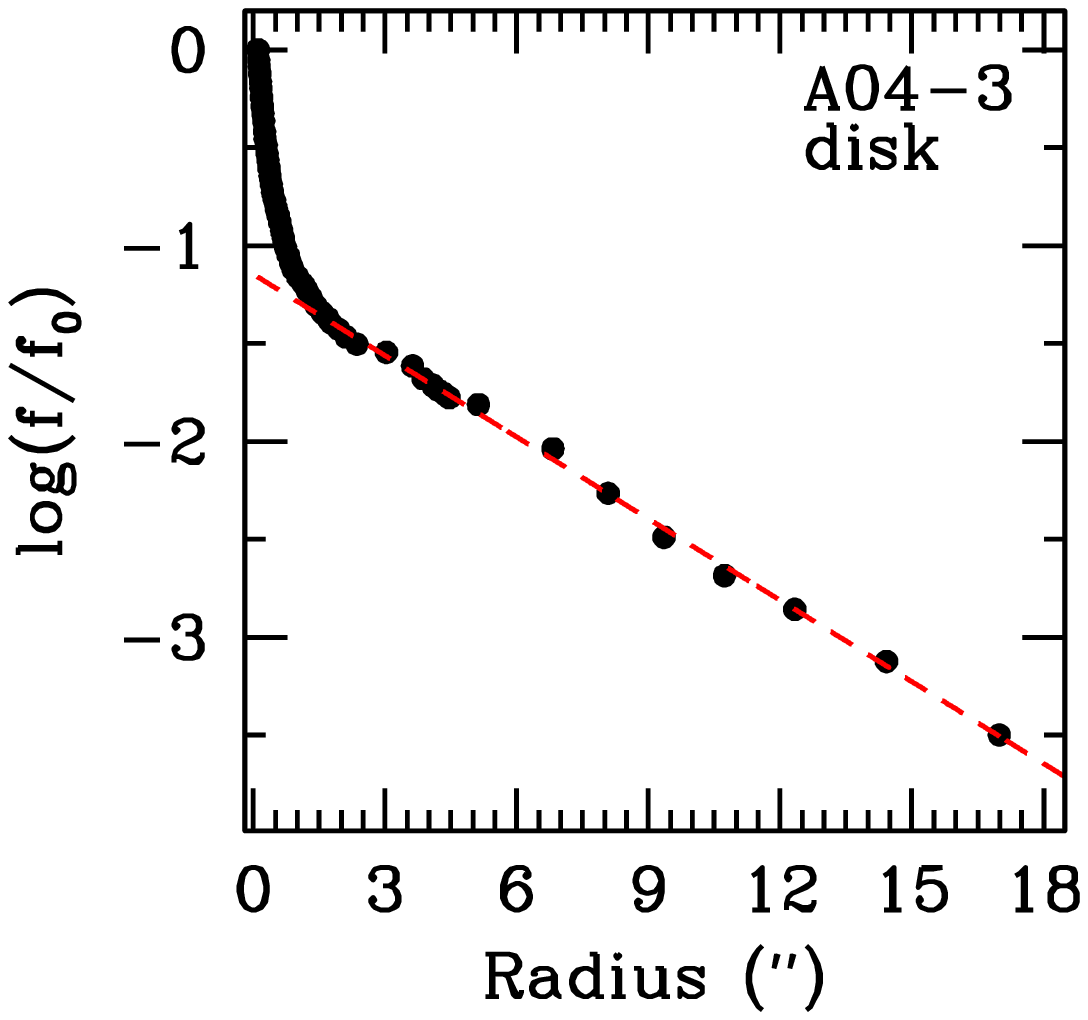}
\end{center}
\caption{Surface brightness profiles of 6500 \AA light for 
  galaxies in the DYNAMO-{\em HST} sample. The galaxies are ordered the
  same as in Fig.~\ref{fig:maps}. Each surface brightness profile is
  normalized by the central surface brightness of the galaxy.  When
  appropriate an exponential disk model that has been fit to the data
  is also shown, represented by a dashed line. The majority of
  DYNAMO-{\em HST} galaxies are consistent both with rotating kinematics
  and, as shown here, an exponential surface photometry of star
  light. Therefore multiple lines of evidence suggest that a disk
  model is appropriate for these galaxies.   }
   \label{fig:profs}
\end{figure*}

As noted earlier, galaxies in the DYNAMO-{\em HST} sample were selected to
have rotating, disk-like kinematics based on the classifications given
in \cite{green2014}. Recently, \cite{bekiaris2015arxiv} fit kinematic
models to many of the galaxies labeled as rotating systems in
\cite{green2014}, and confirmed that they are well-described by models
of disks.  However, in extremely gas rich systems (and many DYNAMO
galaxies are likely to be gas-rich, c.f. \cite{fisher2014}), it is
possible for the gas in galaxy-galaxy mergers to quickly form a disk
\citep{robertson2006}, in which case H$\alpha$ might be a biased
kinematical tracer that is decoupled from the motions of the bulk of
the stars in a galaxy. Though \cite{bassett2014} does confirm that gas
and stars have similar kinematics, on an galaxy-by-galaxy basis there
could be exceptions to this result.  To test whether the DYNAMO-{\em HST}
galaxies are consistent with disk properties, we used data from the
FR647M continuum filter to analyze the surface brightness profiles of
the starlight in sample, using the code and procedure outlined in
\cite{fisherdrory2008}.  Figure~\ref{fig:profs} shows the stellar
surface brightness profiles of all 10 galaxies in the DYNAMO-{\em HST}
sample, and (when appropriate) the corresponding best-fit exponential
disk models.

{\em HST-DYNAMO Disk Galaxies}: Galaxies A04-3, C13-1, D13-5, D15-3,
G04-1, G08-5, G14-1 and G20-2 are consistent with having a disk-like
exponential surface brightness profile of stars. Of these, galaxies
A04-3, C13-1, D13-5, D15-3 and G08-5 have surface photometry that is
consistent with a pure exponential disk, while G04-1, G14-1, and G20-2
are consistent with a disk plus a modest (B/T $<$ 20\%) bulge.
\citep[Note that comparable bulge-to-total ratios are reported for
G20-2 \& G04-1 in][]{obreschkow2015}.

{\em HST-DYNAMO Merging Galaxies}: Similar to the other galaxies we
attempted to fit traditional exponential profiles G13-1 and H10-2,
however the result is an order of magnitude larger residuals from the
fits. Galaxies G13-1 and H10-2 have surface photometry that show
large, non-central, deviations from single exponential disk or
bulge+disk model that is fit to the data.  Though low resolution WiFes
data suggested rotation \citep{green2014}, subsequent analysis with
higher resolution and deep data from GMOS shows that the kinematic
maps of H10-2 are do not have smooth velocity gradients
\citep{bassett2014}, which would suggest they are not rotating
disks. G13-1 has kinematics consistent with a rotating disk model
\citep{bekiaris2015arxiv}. As we state above, simulations make it
clear that when gas fractions are very high, kinematic separation of
galaxies into ``disks'' and ``merging systems'' can sometimes be
misleading \citep{robertson2006}. Also resolution is also very
important, the data used to fit the disk model has $\sim 1-2$~kpc
resolution \citep{green2014}, which may not be sufficient in the case
of G13-1.  In spite of careful pre-selection to isolate rotating disk
systems, contamination of DYNAMO-{\em HST} by mergers is therefore at
the $\sim20\%$ level, and G13-1, and H10-2 will be referred to as
mergers in the rest of this paper. This is comparable to the merger
rate under similar high SFR systems at $z\sim 1-3$
\citep{forsterschreiber2009}.  


\bsp	
\label{lastpage}
\end{document}